\documentclass[preprint,showpacs,preprintnumbers,amssymb,amsmath,nofootinbib]{revtex4}
\usepackage[dvips]{graphicx}
\usepackage{graphicx}
\usepackage{dcolumn}
\usepackage{bm}
\usepackage{axodraw}

\def\lsim{\raise0.3ex\hbox{$\;<$\kern-0.75em\raise-1.1ex\hbox{$\sim\;$}}}
\def\gsim{\raise0.3ex\hbox{$\;>$\kern-0.75em\raise-1.1ex\hbox{$\sim\;$}}}

\newcommand{\be}{\begin{eqnarray}}
\newcommand{\ee}{\end{eqnarray}}
\newcommand{\bea}{\begin{array}}
\newcommand{\eea}{\end{array}}

\renewcommand{\eqref}[1]{Eq.~(\ref{#1})}

\catcode`\@=11
\catcode`\@=12

\begin{document}
\def\descriptionlabel#1{\bf #1\hfill}
\def\description{\list{}{%
\labelwidth=\leftmargin
\advance \labelwidth by -\labelsep
\let \makelabel=\descriptionlabel}}

\title{\vspace{1cm}\Large Like-sign dimuon asymmetry 
of $B^0$ meson and LFV\\
in $SU(5)$ SUSY GUT with $S_4$ flavor symmetry
\vspace{1cm}}

\author{Hajime Ishimori$^a$\footnote{E-mail address: ishimori@muse.sc.niigata-u.ac.jp},~
Yuji Kajiyama$^{b,c}$\footnote{E-mail address: kajiyama@muse.sc.niigata-u.ac.jp},~
Yusuke Shimizu$^a$\footnote{E-mail address: shimizu@muse.sc.niigata-u.ac.jp}
 and Morimitsu Tanimoto$^b$\footnote{E-mail address: tanimoto@muse.sc.niigata-u.ac.jp} }

\affiliation{$^a$Graduate~School~of~Science~and~Technology,~Niigata~University, 
Niigata~950-2181,~Japan  \\
$^b$Department of Physics, Niigata University,~Niigata 950-2181, Japan \\
$^c$National Institute of Chemical Physics and Biophysics, Ravala10, Tallinn, 10143, Estonia\\
}

\begin{abstract}
The like-sign dimuon charge asymmetry of the $B$ meson,
which was reported in the D$\O$ Collaboration, is studied
 in the $SU(5)$ SUSY GUT model with $S_4$ flavor symmetry. 
 Additional CP violating effects from the squark sector 
are discussed in $B_s-\bar B_s$ mixing process. 
The predicted like-sign charge asymmetry is in the 
2$\sigma$ range of the combined result of D$\O$ and CDF measurements. 
Since the SUSY contributions in the quark sector affect to the lepton 
sector because of the $SU(5)$ GUT relation, 
two predictions are given in the leptonic processes: 
(i)~both ${\rm BR}(\mu \to e \gamma)$ and the electron EDM 
are close to the present upper bound,
(ii)~the decay ratios of $\tau$ decays, $\tau \to \mu\gamma$ and 
$\tau \to e \gamma$, are 
related to each other via the Cabibbo angle $\lambda_c$: 
${\rm BR}(\tau \to e\gamma)/{\rm BR}(\tau \to \mu\gamma)\simeq \lambda_c^2$. 
 These are  testable at future experiments. 
\vspace{2cm}
\end{abstract}


\maketitle
\section{Introduction}
The CP violation in the $K$ and $B_d$ mesons has been well explained
within the framework of the standard model (SM).
There is one phase, which is a unique source of the CP violation,
so called Kobayashi-Maskawa (KM) phase \cite{Kobayashi:1973fv},
in the quark sector with three families. 
Until now, the KM phase has successfully described all data related with
the CP violation of $K$ and $B_d$ systems.

 However, there could be new sources of the CP violation if the SM is 
extended to the supersymmetric (SUSY) models. The CP violating phases appear
in  soft scalar mass matrices. These contribute to flavor changing 
neutral  currents (FCNC) with the CP violation.
Therefore, we should examine carefully CP violating phenomena
in the quark sector.

The Tevatron experiments have searched possible effect of the CP violation
in the $B$ meson  system \cite{CDF,Abazov:2010hj}.  Recently, the D$\O$ 
Collaboration reported 
the interesting result of the like-sign dimuon charge asymmetry
$A_{sl}^b({\rm D}\O)=-(9.57 \pm 2.51 \pm 1.46 )\times 10^{-3}$ \cite{Abazov:2010hj}. 
This result is larger than the SM prediction 
$A_{sl}^b({\rm SM})=(-2.3 ^{+0.5}_{-0.6}) \times 10^{-4}$ \cite{Lenz:2006hd}
 at the $3.2 \sigma$ level,
which indicates an anomalous CP violating phase arising in the
$B_s$ meson mixing.

Actually,  new physics  have been discussed
to explain the anomalous CP violation in several approaches.  
 As a possibility, new physics contribute to decay width of the $B_s$
meson \cite{Deshpande:2010hy}-\cite{Bai:2010kf}.
Another possibility is to assume new physics does not give additional
 contribution to the decay width but the $B_s-\overline B_s$ mixing
\cite{Choudhury:2010ya}-\cite{Batell:2010qw}.
This typical model is the general SUSY model with gluino-mediated flavor 
and CP violation \cite{King:2010np, Endo:2010fk, Endo:2010yt, Kubo:2010mh, 
Parry:2010ce, Ko:2010mn, Wang:2011ax}.
Relevant mass insertion (MI) parameters and/or squark mass spectrum can explain
the anomalous CP violation in the $B_s$ system.
Since  the squark flavor mixing is restricted in $K$ and  $B_d$ meson
 systems,  the systematic analyses are necessary to clarify the possible 
effect of  squarks.

In this paper, we study the flavor and CP violation 
 within the framework of  the non-Abelian discrete symmetry 
\cite{Ishimori:2010au}  of 
quark and lepton flavors with SUSY.
Then, the flavor symmetry   controls 
 the  squark and slepton  mass matrices as well as 
 the quark and lepton ones.
For example, the predicted squark  mass matrices reflect
 structures of  the quark  mass matrices.
Therefore, squark mass matrices provide us an important
 test for the flavor symmetry.

The non-Abelian discrete symmetry of flavors
has been studied  intensively in the quark and lepton sectors.
 Actually, the recent neutrino data analyses  
\cite{Schwetz:2008er}-\cite{GonzalezGarcia:2010er} indicate 
the tri-bimaximal mixing  \cite{Harrison:2002er}
-\cite{Harrison:2004uh}, which has been at first understood 
based on the non-Abelian finite group 
$A_4$~\cite{Ma:2001dn,Ma:2002ge,Ma:2004zv,Altarelli:2005yp,Altarelli:2005yx}.
Until now,  much progress has been made in the  theoretical
and phenomenological  analysis of $A_4$ flavor model~\cite{Babu:2002in}-
\cite{Smirnov:2011jv}.

An attractive candidate of the flavor symmetry is the $S_4$ group, 
which was  successful to explain  both  quark and lepton mixing
~\cite{Yamanaka:1981pa}-\cite{Merlo:2011vc}.
Especially, $S_4$ flavor models to unify
 quarks and leptons have been proposed
in the framework of the $SU(5)$ SUSY GUT
~\cite{Ishimori:2008fi,Ishimori:2010xk,Hagedorn:2010th,Ding:2010pc}, 
$SO(10)$  SUSY GUT \cite{Hagedorn:2006ug,Dutta:2009bj,Patel:2010hr}, 
and the Pati-Salam SUSY GUT \cite{Toorop:2010yh,Toorop:2010zg}.
These unified models seem to explain both mixing of quarks and leptons.

Some of us have studied $S_4$ flavor model \cite{Ishimori:2010xk}, 
which gives the proper quark flavor mixing angles
  as well as the tri-bimaximal mixing of neutrino flavors.
 Especially, the Cabibbo angle is predicted to be  $15^\circ$
due to $S_4$ Clebsch-Gordan coefficients.
Including the next-to-leading corrections of the $S_4$ symmetry, the 
predicted   Cabibbo angle is completely consistent with the observed one.
 
We give the squark mass matrices 
in our $S_4$ flavor model
by considering  the gravity mediation within the 
framework of the supergravity theory. 
We estimate the SUSY breaking in the squark mass matrices
by taking account of  the next-to-leading 
$S_4$ invariant mass operators as well as the slepton mass matrices.
Then,  we can predict  the  CP violation in the $B_s$ meson
 taking account of the constraints of the CP violation of  $K$ and $B_d$  
mesons. 
We also discuss the squark effect on $b\rightarrow s\gamma$ decay and 
   the chromo--electric dipole moment (cEDM) .

Since our model is based on $SU(5)$ SUSY GUT, we can predict
 the lepton flavor violation (LFV), e.g., 
$\mu \rightarrow e \gamma$ and  $\tau \rightarrow \mu \gamma$
processes \cite{Ishimori:2010su}. 
In particular,  the $\tau \rightarrow \mu \gamma$ decay ratio
reflects the magnitude of the  CP violation of the $B_s$ meson.

This paper is organized as follows: 
In section 2,
we discuss the possibility of new physics
  in the  framework of  the CP violation of the neutral $B$ system. 
In section 3,
we present  briefly our  $S_4$ flavor model of
 quarks and leptons in  $SU(5)$ SUSY GUT, and present 
 the squark and slepton mass matrices.
In section 4, we discuss numerically  the CP violation of the $B_s$ meson 
 with constraints of flavor and CP violations of $K$ and  $B_d$ mesons.
We also discuss the EDM of the electron, cEDM of strange quark and LFV.
Section 5 is devoted to the summary.
In appendices, we present relevant formulae  in order to estimate
 the flavor violation  and the CP violation.


\section{$ B_s-\bar B_s$ Mixing}
In this section, we briefly discuss the theory and experimental results of  
the CP violation of the neutral $B$ meson system. 
The effective Hamiltonian ${\cal H}^q_{\rm eff}(q=d,s)$ of $B_q- \bar B_q$ system is given 
in terms of the dispersive (absorptive) part $M^q(\Gamma^q)$ as 
\be
{\cal H}^q_{\rm eff}=M^q-\frac{i}{2}\Gamma^q, 
\ee
where the off-diagonal elements $M_{12}^q$ and $\Gamma^q_{12}$ are responsible for 
the $B_q-\bar B_q$ oscillations. The light $(L)$ and heavy $(H)$ physical eigenstates 
$B^q_{L(H)}$ with mass $M^q_{L(H)}$ and the decay 
width $\Gamma^q_{L(H)}$ are obtained by 
diagonalizing the effective Hamiltonian ${\cal H}^q_{\rm eff}$. 
The mass and decay width difference between $B^q_{L}$ and $B^q_H$ 
are related to the elements of ${\cal H}^q_{{\rm eff}}$ as 
\be
\Delta M_q\equiv M^q_H-M^q_L=2|M_{12}^q|,~~~~
\Delta \Gamma_q\equiv \Gamma_L^q-\Gamma_H^q=2 |\Gamma_{12}^q|\cos \phi_q,~~~~
\phi_q={\rm arg}(-M_{12}^q/\Gamma_{12}^q), 
\ee
where we have used $\Delta \Gamma_q \ll \Delta M_q$. 

The ``wrong-sign'' charge asymmetry $a_{sl}^q$ of $B_q \to \mu^- X$ decay is defined as 
\be
a_{sl}^q\equiv \frac{\Gamma(\bar B_q \to \mu^+ X)-\Gamma(B_q \to \mu^- X)}
{\Gamma(\bar B_q \to \mu^+ X)+\Gamma(B_q \to \mu^- X)}
\simeq {\rm Im}\left( \frac{\Gamma_{12}^q}{M_{12}^q}\right)
=\frac{|\Gamma_{12}^q|}{|M_{12}^q|}\sin \phi_q \ . 
\ee
The like-sign dimuon charge asymmetry $A_{sl}^b$ is defined and related with $a_{sl}^q$ as 
\cite{Grossman:2006ce}
\be
A_{sl}^b\equiv \frac{N_b^{++}-N_b^{--}}{N_b^{++}+N_b^{--}}
=(0.506\pm 0.043)a_{sl}^d+(0.494\pm 0.043) a_{sl}^s,
\ee
where $N_b^{\pm \pm}$ is the number of events of $b \bar b\to \mu^{\pm}\mu^{\pm}X$. 

The SM prediction of $A^{b}_{sl}$ is given as \cite{Lenz:2006hd}
\be
A_{sl}^b({\rm SM})=(-2.3 ^{+0.5}_{-0.6}) \times 10^{-4}, 
\label{AbSM}
\ee
which is calculated from \cite{Lenz:2006hd}
\footnote{Recently, the SM predictions are updated \cite{Nierste:2011ti} by the same authors. 
However in this paper, we use the widely-accepted results of Ref. \cite{Lenz:2006hd}.}
\be
a_{sl}^d({\rm SM})=(-4.8^{+1.0}_{-1.2}) \times 10^{-4}, \quad
a_{sl}^s({\rm SM})=(2.06 \pm 0.57) \times 10^{-5}. 
\label{adsSM}
\ee
Recently, the D$\O$ collaboration reported $A_{sl}^b$ with 
6.1 fb$^{-1}$ data set as \cite{Abazov:2010hj} 
\be
A_{sl}^b({\rm D}\O)=-(9.57 \pm 2.51 \pm 1.46 )\times 10^{-3},
\label{AbD0}
\ee
which shows 3.2 $\sigma$ deviation from the SM prediction of Eq.(\ref{AbSM}). 
On the other hand, the result by the CDF collaboration with 1.6 fb$^{-1}$ data  \cite{CDF} 
$A_{sl}^b({\rm CDF})=(8.0 \pm 9.0 \pm 6.8 )\times 10^{-3}$ 
is consistent with the SM prediction while it has large errors. 
Combining these measurements, one can obtain
\be
A_{sl}^b({\rm CDF+D}\O)=-(8.5 \pm 2.8) \times 10^{-3}, 
\label{Abexp}
\ee
which is still 3 $\sigma$ away from the SM prediction. 

The D$\O$ Collaboration have performed the direct measurement of $a_{sl}^s$ 
\cite{Abazov:2009wg} as 
$a_{sl}^s({\rm D}\O)=-(1.7\pm 9.1 ^{+1.4}_{-1.5})\times 10^{-3}$, 
which is consistent with the SM prediction because of its large errors. 
However, if one use the present experimental value of $a_{sl}^d$
 \cite{Abazov:2010hj,Barberio:2008fa,Asner:2010qj}, 
$a_{sl}^d({\rm exp})=-(4.7 \pm 4.6)\times 10^{-3}$, 
one can find that \cite{Abazov:2010hj,Asner:2010qj}
\be
a_{sl}^s=-0.0146 \pm 0.0075, 
\ee
is required to obtain $A_{sl}^b({\rm D}\O)$. The central value of the required $|a_{sl}^s|$ is 
about three orders of magnitude larger than the SM prediction $a_{sl}^s ({\rm SM})$. 
Combining  all results, one can obtain the average value 
\be
a_{sl}^s({\rm average})\simeq -(12.7 \pm 5.0) \times 10^{-3}, 
\ee 
which is still 2.5 $\sigma$ away from the SM prediction $a_{sl}^s ({\rm SM})$. 
Therefore, if the D$\O$ result is confirmed, it is a promising hint of new physics (NP) beyond the SM. 
 
The contribution of NP to the dispersive part of the Hamiltonian 
is parameterized as 
\be
M_{12}^q=M_{12}^{q,SM}+M_{12}^{q,NP}=M_{12}^{q,SM}\left( 1+h_q e^{2 i \sigma_q}\right)
=M_{12}^{q,SM} \Delta_q, \quad
\Delta_q=\left| \Delta_q\right|e^{i \phi_{\Delta q}},
\label{hsigmaDelta}
\ee
where the SM contribution $M_{12}^{q,SM}$ is given by 
\be
M_{12}^{q,SM}=\frac{G_F^2 M_{B_{q}}}{12 \pi^2}M_W^2 (V_{tb}V^*_{tq})^2 \hat \eta_B 
S_0 (x_t) f^2_{B_q}B_q,
\ee
with parameters listed in Table 1.

\begin{table}
\begin{center}
\begin{tabular}{|c|c||c|c|}
\hline
Input &  &  Input &   \\ \hline
$f_{B_s}$ & 
$(231 \pm 3\pm 15)~{\rm MeV}$
& $B_s(m_b)$ 
& $0.841 \pm 0.013 \pm 0.020$
\\ \hline
$f_{B_s}/f_{B_d}$ & $1.209 \pm 0.007 \pm 0.023$ & 
$B_{s}/B_{d} $ & $1.01\pm 0.01 \pm 0.03$
\\ \hline
$\hat \eta_B$ & $0.8393\pm 0.0034$& $S_0(x_t)$ & 
$2.35$
\\ \hline
$M_{s}$ & $5.3663\pm 0.0006$ GeV
 & $\Delta M_{s}^{\rm exp}$ & $17.77\pm 0.10\pm 0.07
 ~\mbox{ps}^{-1}$
\\ \hline
$M_{d}$ & $5.27917\pm 0.00029$ GeV 
& $\Delta M_{d}^{\rm exp}$ & 
$0.507\pm 0.005 ~\mbox{ps}^{-1}$ 
\\ \hline
$m_d(m_b)$ & $(5.1\pm 1.3)
\times 10^{-3}$ GeV & $m_s (m_b)$ & $0.085\pm 0.017$ GeV
\\ \hline
$m_b(m_b)$ & $4.248\pm 0.051$ GeV&$\Delta \Gamma_s ^{SM}$&$(0.096 \pm 0.039)~{\rm ps}^{-1}$
\\ \hline
$\phi_{d,SM}$&$\left( -10.1^{+3.7}_{-6.3}\right) \times 10^{-2}$&
$\phi_{s,SM}$&$\left( +7.4^{+0.8}_{-3.2}\right) \times 10^{-3}$
\\ \hline
$| \Gamma_{12}^{s,SM} |/| M_{12}^{s,SM}|$&$(4.97 \pm 0.94)\times 10^{-3}$&
$\Delta \Gamma_d^{SM}/\Delta M_d^{SM}$&$(52.6 ^{+11.5}_{-12.8})\times 10^{-4}$
\\ \hline
\end{tabular}
\caption{Parameters of the neutral $B$ meson mixing and quark masses
\cite{Lenz:2006hd,Lenz:2010gu}. }
\label{table1}
\end{center}
\end{table}

Using these parameters, the mass difference of 
$B_q$ meson, $\Delta M_q$, is given by 
\be
\Delta M_q=\Delta M_q^{SM}\left|1+h_q e^{2 i \sigma_q} \right|
=\Delta M_q^{SM}\left| \Delta_q\right|.
\ee
Since the SM contribution to the absorptive part $\Gamma_{12}^s$ 
is dominated by tree-level decay $b\to c \bar c s$, 
one can set $\Gamma_{12}^s=\Gamma_{12}^{s,SM}$. 
In this case, the wrong-sign charge asymmetry $a_{sl}^s$ is 
written as
\be
a_{sl}^q
=\frac{| \Gamma_{12}^{q,SM} |}{| M_{12}^{q,SM}|}
\frac{\sin \left( \phi_{q,SM}+\phi_{\Delta q}\right)}{\left|\Delta_q\right|}. 
\label{aslq}
\ee
Taking the experimental value $\Delta M_s^{\rm exp}$  
into account, one finds that 
$|\Delta_s|$ is strongly constrained in the region $|\Delta_s|=0.92 \pm 0.32$ \cite{Lenz:2006hd}. 
Therefore, unphysical condition $\sin (\phi_{s,SM}+\phi_{\Delta s})=-2.56 \pm 1.16$ is required to obtain 1 $\sigma$ range of the charge asymmetry (See also \cite{Berger:2010wt}). 
Also as discussed in Ref \cite{Chen:2010aq}, by using the SM prediction of $\Gamma^{d,s}_{12}$ 
and experimental values of $\Delta M_{d,s}$, they found  in model-independent way 
that the like-sign charge asymmetry is bounded as 
$-A_{sl}^b<3.16 \times 10^{-3}$,  
where the CP violation $S_{J/\psi K_S}$ and $S_{J/\psi \phi}$ are also taken into account.  

Now we discuss how to avoid this unphysical condition to obtain large 
charge asymmetry. 
As the first possibility, one can consider the NP contributions to 
$\Gamma_{12}^s$, which come from additional contributions to decay processes
$b \to c \bar c s,~\tau^+ \tau^- s$, etc.
By using  the D$\O$ and CDF experimental data  of $B_s \to J/\psi \phi$ 
decay \cite{CDFD02009}, 
one can subtract $\Delta \Gamma_s$ and $\beta_s^{J/\psi \phi}\simeq-\phi_s/2$ as \cite{Barberio:2008fa} 
\footnote{See also Ref.\cite{Asner:2010qj} for recent results.}
\be
\Delta \Gamma_s=\pm (0.154 ^{+0.054}_{-0.070})~{\rm ps}^{-1},\qquad
\beta_s^{J/\psi \phi}=(0.39^{+ 0.18}_{-0.14})~{\rm or}~ (1.18^{+0.14}_{-0.18}), 
\label{gammabetaexp}
\ee 
where the sign of $\Delta \Gamma_s$ is still undetermined, and positive (negative) sign 
corresponds to the first (second) region of $\beta_s^{J/\psi \phi}$. 
Comparing them with the SM predictions, one finds that there still can 
exist additional contributions to $\Gamma_{12}^s$ 
from NP. This possibility has been studied in several models\footnote{However, the NP contributions to 
$\Gamma _{12}^q$ will be strongly constrained by the lifetime ratio $\tau _{B_s}/\tau _{B_d}$. 
We would like to thank A. Lenz for pointing out this point.} 
\cite{Deshpande:2010hy,Oh:2010vc,He:2010fz,Dighe:2010nj,
Bauer:2010dga,Chao:2010mq,Datta:2010yq,Bai:2010kf}. 

In Ref.\cite{Choudhury:2010ya}, while there are no NP contributions to 
$\Gamma_{12}^s$ in their model, they employed the experimental value of 
$\Delta \Gamma_s$ of Eq.(\ref{gammabetaexp}) since there must exist 
theoretical uncertainties. In the $(h_q,\sigma_q)$ parametrization 
of NP, the best fit values of $(h_s,\sigma_s)$ are obtained as
\cite{Ligeti:2010ia}
\be
(h_s,\sigma_s)\simeq (0.5,120^\circ),~(1.8,100^\circ), 
\label{bestfit}
\ee
by taking $\Delta M_q,~ \Delta \Gamma_q,~S_{\psi K}$ and $S_{J/\psi \phi}$ 
into account, 
with varying $|\Gamma_{12}^s|$ in the range $0-0.3{\rm ps}^{-1}$. 
In  that paper, one can read that the region of $h_d \lsim h_s$ is favored
as seen in  Refs.  \cite{King:2010np,Endo:2010fk,Endo:2010yt}. 

However in ordinary SUSY models, gluino-squark box diagrams do not give 
additional contributions to $\Gamma_{12}^s$ since such diagrams do not 
generate additional decay modes of bottom quark.  
Therefore as the other possibility, constraint for $\Delta M_q$ is 
relaxed in Refs.\cite{Kubo:2010mh,Parry:2010ce}.  
In those papers, they consider models that NP does not give additional 
contributions to 
$\Gamma_{12}^s$, but to $M_{12}^q$. They take a conservative constraint 
$0.6< \Delta M_{d,s}/\Delta M_{d,s}^{\rm exp}<1.4$ \cite{Kubo:2010mh} and 
the UTfit \cite{Bona:2008jn} allowed region 
$0.776< \Delta M_{d,s}/\Delta M_{d,s}^{SM}<1.162$ \cite{Parry:2010ce}. 
See also Refs. \cite{Chen:2010aq,Dobrescu:2010rh,Ko:2010mn,
Wang:2011ax,Park:2010sg,Kostelecky:2010bk} for other possibilities.

In this paper, we consider the NP contribution to $B_s-\bar B_s$ mixing by 
gluino-squark box diagrams in a $SU(5)$ SUSY GUT model with $S_4$ flavor symmetry. 
As shall be discussed in the next section, the soft SUSY breaking terms and related MI parameters 
$(\delta_d^{AB})_{ij}(A,B=L,R)$ obey $S_4$ flavor symmetry. In such SUSY models, 
there are no new contributions to $\Gamma_{12}^s$
\cite{King:2010np, Endo:2010fk, Endo:2010yt, Kubo:2010mh, Parry:2010ce, 
Ko:2010mn, Wang:2011ax}. 
While the SUSY contributions to $B_s-\bar B_s$ mixing are 
induced by $(\delta_d^{AB})_{23}$, it is constrained by $b \to s\gamma$ decay. 
Since the other MI parameters of down-type squark sector are related to 
$(\delta_d^{AB})_{23}$ due to $S_4$ symmetry, 
$K$ and $B_d$ meson mixing, 
which are affected by $(\delta_d^{AB})_{12}$ and $(\delta_d^{AB})_{13}$, 
respectively, should also be taken into account. 
The CP violation in $B_s$ meson system is related to 
cEDM of the strange quark $d^C_s$ as well. 
Moreover, the leptonic processes such as $\tau \to \mu \gamma$ affected by 
$(\delta_{\ell}^{AB})_{23}$ should also be 
taken into account due to $SU(5)$ GUT relation. 

Taking the above processes into account, we assume the following conditions in our 
numerical calculation:  
(i) the meson mass differences satisfy
\be
0.6< \frac{\Delta M_{d,s}}{\Delta M_{d,s}^{\rm exp}}<1.4,\qquad
\frac{|M_{12}^{K,SUSY}|}{\Delta M_K^{\rm exp}}<1,\qquad
\frac{|{\rm Im}M_{12}^{K,SUSY}|}{\sqrt{2}\Delta M_K^{\rm exp}}<\epsilon_K=2.2 \times 10^{-3}, 
\ee 
(ii) cEDM of the strange quark is constrained by the neutron EDM as 
\cite{Hisano:2003iw,Baker:2006ts}
\be
|ed^C_s|<1.0 \times 10^{-25}~e{\rm cm}, 
\ee
(iii) the NP contribution to the branching ratio (BR) of $b \to s \gamma$ is constrained as 
\be
{\rm BR}(b\to s \gamma)^{NP}&<&1.0 \times 10^{-4}. 
\ee
While the upper bounds of LFV decay processes $\ell_i \to \ell_j \gamma$ and the 
electron EDM are given by \cite{Nakamura:2010zzi,Altmannshofer}
\be
 {\rm BR}(\mu\to e \gamma)&<&1.2 \times 10^{-11},\qquad
  {\rm BR}(\tau\to \mu \gamma)<4.4 \times 10^{-8},~~\\
   {\rm BR}(\tau\to e \gamma)&<&3.3 \times 10^{-8}, \qquad
   |ed_e|<1.6 \times 10^{-27}~e{\rm cm}, 
   \label{lfvbound}
\ee
we do not take these bounds into account in the numerical calculation below. 
Instead, in the allowed parameter region of our model which can explain the like-sign charge asymmetry, 
we will obtain the predictions for LFV processes.  

We perform numerical analysis in the section IV after introducing the $S_4$ flavor model 
in the next section. 


\section{The $S_4$ flavor model}
\label{sec:model}

We briefly review $S_4$ flavor model of quarks and leptons,
which was proposed in \cite{Ishimori:2010xk}. 
As the model is based on $SU(5)$ SUSY GUT, 
it gives sfermion mass matrices as well as quark and lepton mass matrices.

\vspace{0.5cm}
\begin{table}[t]
\begin{tabular}{|c|ccccc||cccc|}
\hline
&$(T_1,T_2)$ & $T_3$ & $( F_1, F_2, F_3)$ & $(N_e^c,N_\mu ^c)$ & $N_\tau ^c$ & $H_5$ &$H_{\bar 5} $ & $H_{45}$ & $\Theta $ \\ \hline
$SU(5)$ & $10$ & $10$ & $\bar 5$ & $1$ & $1$ & $5$ & $\bar 5$ & $45$ & $1$ \\
$S_4$ & $\bf 2$ & $\bf 1$ & $\bf 3$ & $\bf 2$ & ${\bf 1}'$ & $\bf 1$ & $\bf 1$ & $\bf 1$ & $\bf 1$ \\
$Z_4$ & $-i$ & $-1$ & $i$ & $1$ & $1$ & $1$ & $1$ & $-1$ & $1$ \\
$U(1)_{FN}$ & $1 $ & 0 & 0 & $1$ & 0 & 0 & 0 & 0 & $-1$ \\
\hline
\end{tabular}
\vspace{+0.5cm}
\begin{tabular}{|c|cccccc|}
\hline
& $(\chi _1,\chi _2)$ & $(\chi _3,\chi _4)$ & $(\chi _5,\chi _6,\chi _7)$ 
& $(\chi _8,\chi _9,\chi _{10})$ & $(\chi _{11},\chi _{12},\chi _{13})$ & $\chi _{14}$ \\ \hline
$SU(5)$ & $1$ & $1$ & $1$ & $1$ & $1$ & $1$  \\
$S_4$ & $\bf 2$ & $\bf 2$ & ${\bf 3}'$ & $\bf 3$ & $\bf 3$ & $\bf 1$ \\
$Z_4$ & $-i$ & $1$ & $-i$ & $-1$ & $i$ & $i$\\
$U(1)_{FN}$ & $-1 $ & $-2$ & 0 & 0 & 0 & $-1 $ \\
\hline
\end{tabular}
\caption{Assignments of $SU(5)$, $S_4$, $Z_4$, and $U(1)_{FN}$ representations.}
\label{tables4}
\end{table}

\subsection{CKM Mixing}
In the $SU(5)$ GUT, matter fields are unified into $10$ 
and $\bar 5$-dimensional representations as $ 10 \subset (Q,u^c,e^c)$ and 
$\bar 5\subset (d^c,L)$. 
Three generations of $ \bar 5$, which are denoted by $F_i~(i=1,2,3)$,
 are assigned to $\bf 3$ of $S_4$. 
On the other hand, the third generation of the $ 10$-dimensional 
representation, $T_3$, is assigned to $\bf 1$ of $S_4$, and  
 the first and second generations of $10$, $(T_1, T_2)$, 
are assigned to $\bf 2$ of $S_4$, respectively. 
Right-handed neutrinos, which are $SU(5)$ gauge singlets, 
are also assigned to ${\bf 2}$ for the first and second generations,
$(N_e^c, N_\mu^c)$,
and ${\bf 1}'$ for  the third one, $N_\tau^c$. 
The $5$-dimensional, 
$\bar 5$-dimensional,  and $45$-dimensional Higgs of $SU(5)$, $H_5$, 
$H_{\bar 5} $, and $H_{45} $ are  assigned to $\bf 1$ of $S_4$. 
In order to obtain desired mass matrices, we introduce 
 $SU(5)$ gauge singlets $\chi_i$, so called flavons,  
which couple to quarks and leptons.


The $Z_4$ symmetry is added to obtain relevant couplings.
Further, the Froggatt-Nielsen mechanism~\cite{Froggatt:1978nt}
  is  introduced  to get the natural hierarchy among quark and lepton masses,
as an additional 
$U(1)_{FN}$ flavor symmetry, where 
$\Theta$ denotes the Froggatt-Nielsen flavon.
The particle assignments of $SU(5)$, $S_4$, $Z_4$, and $U(1)_{FN}$
 are presented  in Table \ref{tables4}. 

The couplings of flavons with fermions are restricted as follows.
At the leading order, $(\chi _3,\chi _4)$ are  
coupled with the right-handed Majorana neutrino sector, 
$(\chi _5,\chi _6,\chi _7)$ are coupled with the Dirac neutrino sector, 
$(\chi _8,\chi _9,\chi _{10})$ and $(\chi _{11},\chi _{12},\chi _{13})$ 
are coupled with the charged lepton and down-type quark sectors.
At the next-to-leading order, 
$(\chi_1,\chi_2)$ are coupled with the  up-type  quark sector, 
and  $\chi_{14}$ contributes 
 to the charged lepton and down-type quark sectors,
 and then the mass ratio of the electron and down quark is reproduced
properly.

Our model predicts the quark  mixing  as well as the tri-bimaximal
mixing of leptons. Especially, the Cabibbo angle is
predicted to be     $15^{\circ}$ at the leading order.
The model is consistent with   the observed CKM mixing angles
and CP violation
 as well as the non-vanishing $U_{e3}$ of the neutrino flavor mixing.


 Let us write down  the superpotential 
respecting  $S_4$, $Z_4$ and $U(1)_{FN}$
symmetries
 in terms of the $S_4$ cutoff scale $\Lambda$,  and
the $U(1)_{FN}$ cutoff scale  $\overline \Lambda$. 
In our calculation, 
both cutoff scales are taken as the GUT scale which is around $10^{16}$GeV. 
The $SU(5)$ invariant superpotential 
of the Yukawa  sector up to the linear terms of $\chi_i$ ($i=1,\cdots ,13$) 
is given as
\begin{align}
w &= y_1^u(T_1,T_2)\otimes T_3\otimes (\chi _1,\chi _2)\otimes H_5/\Lambda + y_2^uT_3\otimes T_3\otimes H_5 \nonumber \\
&\ + y_1^N(N_e^c,N_\mu ^c)\otimes (N_e^c,N_\mu ^c)\otimes \Theta ^{2}/\bar \Lambda  \nonumber \\
&\ + y_2^N(N_e^c,N_\mu ^c)\otimes (N_e^c,N_\mu ^c)\otimes (\chi _3,\chi _4) + MN_\tau ^c\otimes N_\tau ^c \nonumber \\
&\ + y_1^D(N_e^c,N_\mu ^c)\otimes (F_1,F_2,F_3)\otimes (\chi _5,\chi _6,\chi _7)\otimes H_5\otimes \Theta /(\Lambda \bar \Lambda ) \nonumber \\
&\ + y_2^DN_\tau ^c\otimes (F_1,F_2,F_3)\otimes (\chi _5,\chi _6,\chi _7)\otimes H_5/\Lambda \nonumber \\
&\ + y_1(F_1,F_2,F_3)\otimes (T_1,T_2)\otimes (\chi _8,\chi _9,\chi _{10})\otimes H_{45}\otimes \Theta /(\Lambda \bar \Lambda) \nonumber \\
&\ + y_2(F_1,F_2,F_3)\otimes T_3\otimes (\chi _{11},\chi _{12},\chi _{13})\otimes H_{\bar 5}/\Lambda ,
\end{align}
where $y_1^u$, $y_2^u$, $y_1^N$, $y_2^N$, $y_1^D$, $y_2^D$, 
$y_1$, and $y_2$ are Yukawa couplings of order one, 
and $M$ is the right-handed Majorana mass, which is taken to be $10^{12}$GeV. 

 In order to predict the desired quark and lepton mass matrices,
 we require vacuum alignments for the vacuum expectation values (VEV's) 
of flavons. According to the potential analysis, which was presented
 in  \cite{Ishimori:2010xk}, 
we have conditions of VEV's to realize  the potential minimum ($V=0$) 
as follows:
\begin{eqnarray}
&& (\chi _1, \chi _2)= (1,1), \quad 
 (\chi _3, \chi _4)= (0,1), \quad
 (\chi _5, \chi _6, \chi _7)=(1,1,1),
 \quad (\chi _8, \chi _9, \chi _{10})=(0,1,0), 
\nonumber\\
&&(\chi _{11}, \chi_{12}, \chi _{13})=(0,0,1), 
\quad \chi_{14}^2=-\frac{2\eta_2}{\eta_{3}}\chi_1^2,
\label{alignment}
\label{vev}
\end{eqnarray}
where  these magnitudes are given in  arbitrary units.
Hereafter, we suppose these gauge-singlet scalars 
develop VEV's by denoting $\langle \chi_i\rangle=a_i\Lambda$.  

Denoting Higgs doublets as $h_u$
and $h_d$, we take VEV's of following scalars by
\begin{eqnarray}
\langle h_u\rangle 
= v_u, 
\quad
\langle h_d\rangle 
= v_d, 
\quad
\langle h_{45}\rangle 
= v_{45}, 
\quad 
\langle\Theta\rangle =\theta ,
\end{eqnarray}
which are supposed to be real. 
We define $\lambda \equiv \theta/\Lambda$
to describe the Froggatt-Nielsen mechanism
\footnote{Notice that this $\lambda$ is not related to the Cabibbo angle 
$\lambda_c$ in our model. }.

Now, we can write down quark and lepton mass matrices 
by using the $S_4$ multiplication rule in Appendix A.
The down-type quark mass matrix at the leading order is given as
\begin{equation}
M_d = \begin{pmatrix}
            0 & 0 & 0 \\ 
            y_1\lambda a_9 v_{45}/\sqrt 2 & y_1\lambda
 a_9 v_{45}/\sqrt 6 & 0 \\
            0 & 0 & y_2a_{13}v_d
         \end{pmatrix}. 
\end{equation}
Then, we have
\begin{equation}
M_d^\dagger M_d = v_d^2\begin{pmatrix}
                          \frac{1}{2}|\bar y_1\lambda a_9|^2 & \frac{1}{2\sqrt 3}|\bar y_1\lambda a_9|^2 & 0 \\
                          \frac{1}{2\sqrt 3}|\bar y_1\lambda a_9|^2 & \frac{1}{6}|\bar y_1\lambda a_9|^2 & 0 \\
                          0 & 0 & |y_2|^2a_{13}^2
                       \end{pmatrix},
\end{equation}
where we denote $\bar y_1v_d=y_1v_{45}$.
This matrix can be diagonalized  by the orthogonal matrix $U_d^{(0)}$ as
\begin{align}
U_d^{(0)} = \begin{pmatrix}
            \cos 60^\circ & \sin 60^\circ & 0 \\
            -\sin 60^\circ & \cos 60^\circ & 0 \\
            0 & 0 & 1
         \end{pmatrix}.
\label{Ud}
\end{align}
The down-type quark masses  are given as 
\begin{align}
&m_d^2=0\ ,
\quad  
m_s^2=\frac{2}{3}|\bar y_1\lambda a_9|^2v_d^2\ ,
\quad 
m_b^2=|y_2|^2a_{13}^2v_d^2\ .
\label{downmass}
\end{align}
The down quark mass vanishes, 
however tiny masses appear at  the next-to-leading order. 

The relevant superpotential of down sector 
at the next-to-leading order  is given as 
\begin{align}
\Delta w_d&=y_{\Delta _a}(T_1,T_2)\otimes (F_1,F_2,F_3)\otimes (\chi _1,\chi _2)\otimes (\chi _{11},\chi _{12},\chi _{13})\otimes H_{\bar 5}/\Lambda ^2 \nonumber \\
&\ +y_{\Delta _b}(T_1,T_2)\otimes (F_1,F_2,F_3)\otimes (\chi _5,\chi _6,\chi _7)\otimes \chi _{14}\otimes H_{\bar 5}/\Lambda ^2 \nonumber \\
&\ +y_{\Delta _c}(T_1,T_2)\otimes (F_1,F_2,F_3)\otimes (\chi _1,\chi _2)\otimes (\chi _5,\chi _6,\chi _7)\otimes H_{45}/\Lambda ^2 \nonumber \\
&\ +y_{\Delta _d}(T_1,T_2)\otimes (F_1,F_2,F_3)\otimes (\chi _{11},\chi _{12},\chi _{13})\otimes \chi _{14}\otimes H_{45}/\Lambda ^2 \nonumber \\
&\ +y_{\Delta _e}T_3\otimes (F_1,F_2,F_3)\otimes (\chi _5,\chi _6,\chi _7)\otimes (\chi _8,\chi _9,\chi _{10})\otimes H_{\bar 5}\otimes /\Lambda ^2 \nonumber \\
&\ +y_{\Delta _f}T_3\otimes (F_1,F_2,F_3)\otimes (\chi _8,\chi _9,\chi _{10})\otimes (\chi _{11},\chi _{12},\chi _{13})\otimes H_{45}\otimes /\Lambda ^2\ ,
\label{nextsusy}
\end{align} 
which gives the correction terms in the down-type quark mass matrix.

The down-type quark mass matrix including the next-to-leading order is
\begin{equation}
M_d\simeq
\begin{pmatrix}
\bar\epsilon _{11} & \bar\epsilon _{21} & \bar\epsilon _{31} \\
\frac{\sqrt 3m_s}{ 2}+\bar\epsilon_{12} & \frac{m_s}{2}+\bar\epsilon_{22} 
& \bar\epsilon _{32} \\
\bar\epsilon _{13} & \bar\epsilon _{23} & m_b+\bar\epsilon_{33}
\end{pmatrix},
\label{nextleading}
\end{equation}
where the explicit forms of 
$\bar \epsilon_{ij}$'s are given  in Appendix B, and 
 $m_s$ and $m_b$ are given in  Eq.~(\ref{downmass}).
This mass matrix can be diagonalized by
\begin{eqnarray}
M_d^{\rm diag}
=V_d^\dagger M_d U_d^{(0)} U_d^{(1)},
\end{eqnarray}
where mixing matrices for left-hand $U_d^{(1)}$
and for right-hand $V_d$ are estimated as
\begin{eqnarray}
\begin{split}
U_d^{(1)}&=
\begin{pmatrix}
1 & \theta_{12}^d & \theta_{13}^d \\ 
-\theta _{12}^d-\theta _{13}^d\theta _{23}^d  & 1 & \theta _{23}^d \\
-\theta _{13}^d+\theta _{12}^d\theta _{23}^d  & -\theta _{23}^d-\theta _{12}^d\theta _{13}^d & 1 \\
\end{pmatrix},
\\
V_d&=
 \begin{pmatrix}
       1 &  \frac{\tilde a}{\lambda} &  \tilde a\\
    -\frac{\tilde a}{\lambda}-\tilde a^2  & 1  &  \tilde a\\
    -\tilde a+ \frac{\tilde a^2}{\lambda} 
& -\tilde a-  \frac{\tilde a^2}{\lambda}& 1
\end{pmatrix},
\end{split}
\end{eqnarray}
 where $\tilde a$ denotes the typical value of 
 the square root of the relevant sum  of $a_i a_j$'s
 as discussed in the next subsection. 
We neglect CP violating phases then 
mixing angles $\theta _{12}^d,\ \theta _{13}^d,\ \theta _{23}^d$ 
are given as
\begin{eqnarray}
\theta _{12}^d=\mathcal{O}\left (\frac{m_d}{m_s}\right )
=\mathcal{O}\left (0.05\right ),\ \  
\theta _{13}^d=\mathcal{O}\left (\frac{m_d}{m_b}\right )
=\mathcal{O}\left (0.005\right ),\ \ 
\theta _{23}^d=\mathcal{O}\left (\frac{m_d}{m_b}\right )
=\mathcal{O}\left (0.005\right ).
\label{dangles}
\end{eqnarray}


On the other hand, 
the superpotential of up sector 
at the next-to-leading order is 
\begin{align}
\Delta w_u&=y_{\Delta _a}^u(T_1,T_2)\otimes (T_1,T_2)\otimes (\chi _1,\chi _2)\otimes (\chi _1,\chi _2)\otimes H_{5}/\Lambda ^2 \nonumber \\
&\ +y_{\Delta _b}^u(T_1,T_2)\otimes (T_1,T_2)\otimes \chi_{14}\otimes \chi_{14}\otimes H_{5}/\Lambda ^2 \nonumber \\
&\ +y_{\Delta _c}^uT_3 \otimes T_3\otimes (\chi _8,\chi _9,\chi _{10})\otimes (\chi _8,\chi _9,\chi _{10})\otimes H_{5}/\Lambda ^2  .
\end{align}
Then the mass matrix becomes
\begin{equation}
 M_u=v_u
\begin{pmatrix}
2y_{\Delta _{a_1}}^ua _1^2+y_{\Delta _{b}}^ua_{14}^2 & y_{\Delta _{a_2}}^ua _1^2 & y_1^ua _1 \\
y_{\Delta _{a_2}}^ua _1^2 & 2y_{\Delta _{a_1}}^ua _1^2+y_{\Delta _{b}}^ua_{14}^2 & y_1^ua _1 \\
y_1^ua _1 & y_1^ua _1 & y_2^u+y_{\Delta _c}^ua _9^2
\end{pmatrix}.
\end{equation}
This symmetric mass matrix is diagonalized by 
the unitary matrix  $U_u$ as
\begin{eqnarray}
U_u=
\begin{pmatrix}
\cos 45^\circ & \sin 45^\circ & 0 \\
-\sin 45^\circ & \cos 45^\circ & 0 \\
0 & 0 & 1
\end{pmatrix} 
\begin{pmatrix}     
1 & 0 & 0 \\
0 & r_t & r_c \\
0 & -r_c & r_t
\end{pmatrix},
\end{eqnarray}
where  $r_c=\sqrt{m_c/(m_c+m_t)}$ and  $r_t=\sqrt{m_t/(m_c+m_t)}$.

Therefore, the CKM matrix $V$can be written 
as\footnote{The renormalization group effect for 
the CKM matrix is small 
so that the matrix given in the text can be regarded as 
the one at the electroweak scale.}
\begin{equation}
V= U_u^\dagger 
\begin{pmatrix}     
1 & 0 & 0 \\
0 & e^{-i \rho} & 0 \\
0 & 0 & 1
\end{pmatrix}U_d^{(0)} U_d^{(1)}\ ,
\end{equation}
where the phase $\rho$ is an arbitrary parameter originating from complex
Yukawa couplings. 

At the leading order, the Cabibbo angle is derived as 
$60^\circ - 45^\circ=15^\circ$ 
and it can be naturally fitted to the observed value by including
the next-to-leading  order as follows:
\begin{eqnarray}
V_{us}
&\simeq \theta _{12}^d\cos 15^\circ +\sin 15^\circ.
\end{eqnarray}
The $V_{cb}$ and  $V_{ub}$ mixing elements are  expressed  as
\begin{equation}
\begin{split}
V_{ud}
&\simeq \cos 15^\circ -(\theta _{12}^d+\theta _{13}^d\theta _{23}^d)\sin 15^\circ ,
\\
V_{cb}
&\simeq -r_t\theta _{13}^de^{i\rho }\sin 15^\circ +r_t\theta _{23}^de^{i\rho }\cos 15^\circ -r_c\ , \\
V_{ub}
&\simeq \theta _{13}^d\cos 15^\circ +\theta _{23}^d\sin 15^\circ ,
\end{split}
\end{equation}
which are consistent with observed values.

We can also predict  mass matrices of
the charged leptons and neutrinos,
which give   the tri-bimaximal mixing of leptons.
 Details are shown  in Appendix C.


Since mass eigenvalues of quarks and leptons are give in terms of
  $a_i = \langle \chi_i \rangle / \Lambda$, we can estimate
$a_i$ by putting the observed quark and lepton masses.
These are given as
\begin{align}
&a_ 3=a_8=a_{10}=a_{11}=a_{12}=0, 
\qquad 
a_1=a_2\simeq\sqrt{\frac{m_c}{2\left |y_{\Delta _{a_2}}^u-\frac{{y_1^u}^2}{y_2^u}\right |v_u}}~, 
\nonumber\\
&a_4 = \frac{(y_1^D\lambda )^2(m_{\nu_3}-m_{\nu_1})m_{\nu_2}M}
{6y_2^N{y_2^D}^2m_{\nu_1}m_{\nu_3}\Lambda },
\qquad
a _5 =a_6 =a_7 = \frac{\sqrt{m_{\nu_2}M}}{\sqrt 3y_2^Dv_u}, 
\nonumber \\
& a_9= \frac{m_\mu }{\sqrt 6|\bar{y_1}|\lambda v_d}, 
\qquad a_{13} = \frac{m_\tau }{y_2v_d}\ ,
\label{alphas}
\end{align}
where  masses of quarks and leptons are given at the GUT scale, and 
the light neutrino masses $m_{\nu_{1,2,3}}$ are given in the Appendix C.
Hereafter, we take $\lambda=0.1$ in our calculations.

\subsection{Squark and slepton mass matrices}

Here we study SUSY breaking terms
in the framework of $S_4 \times Z_4 \times U(1)_{FN}$
to derive squark and slepton mass matrices.
We consider the gravity mediation within the 
framework of the supergravity theory.
We assume that 
non-vanishing $F$-terms of gauge and flavor singlet (moduli) fields $Z$ 
and gauge singlet fields $\chi_i$ $(i=1,\cdots,14)$ 
contribute to the SUSY breaking.
Their $F$-components are written as 
\begin{eqnarray}
F^{\Phi_k}= - e^{ \frac{K}{2M_p^2} } K^{\Phi_k \bar{I} } \left(
  \partial_{\bar{I}} \bar{W} + \frac{K_{\bar{I}}} {M_p^2} \bar{W} \right) ,
\label{eq:F-component}
\end{eqnarray}
where $M_p$ is the Planck mass, $W$ is the superpotential, 
$K$ denotes the K\"ahler potential, $K_{\bar{I}J}$ denotes 
second derivatives by fields, 
i.e. $K_{\bar{I}J}={\partial}_{\bar{I}} \partial_J K$
and $K^{\bar{I}J}$ is its inverse. 
Here the fields ${\Phi_k}$ correspond to the moduli fields $Z$ and 
gauge singlet fields $\chi_i$.
The VEV's of $F_{\Phi_k}/\Phi_k$  are estimated as 
$\langle F_{\Phi_k}/ \Phi_k \rangle = {\cal O}(m_{3/2})$, where
$m_{3/2}$ denotes the gravitino mass, which is obtained as 
$m_{3/2}= \langle e^{K/2M_p^2}W/M_p^2 \rangle$.

First, let us study soft scalar masses.
Within the framework of the supergravity theory,
the soft scalar mass squared is obtained as~\cite{Kaplunovsky:1993rd}
\begin{eqnarray}
m^2_{\bar{I}J} K_{{\bar{I}J}}= m_{3/2}^2K_{{\bar{I}J}} 
+ |F^{\Phi_k}|^2 \partial_{\Phi_k}  
\partial_{  \bar{\Phi_k} }  K_{\bar{I}J}-
|F^{\Phi_k}|^2 \partial_{\bar{\Phi_k}}  K_{\bar{I}L} \partial_{\Phi_k}  
K_{\bar{M} J} K^{L \bar{M}}.
\label{eq:scalar}
\end{eqnarray}
The invariance under the  $S_4 \times Z_4 \times U(1)_{FN}$
flavor symmetry 
as well as the gauge invariance requires the following form 
of the K\"ahler potential as 
\begin{equation}
K = Z^{(5)}(\Phi )\sum_{i=1,2,3 } |F_i|^2 + 
Z_{(1)}^{(10)}(\Phi )\sum_{i=1,2 }|T_i|^2 +Z_{(2)}^{(10)}(\Phi )|R_\tau |^2, 
\label{eq:Kahler}
\end{equation}
at the lowest level, where $Z^{(5)}(\Phi)$ and $Z_{(1),(2)}^{(10)}(\Phi)$ are 
arbitrary functions of the singlet fields $\Phi$.
By use of Eq.~(\ref{eq:scalar}) with 
the K\"ahler potential in Eq.~(\ref{eq:Kahler}), 
we obtain the following matrix form 
of soft scalar masses squared for $\overline {5}\ \overline {5}^c$  and 
 ${10} \ {10}^c$ combinations, which are denoted
as $m_{F}^2$ and $m_{T}^2$, respectively:
\begin{eqnarray}
(m_{F}^2)_{ij} =
\left(
  \begin{array}{ccc}
m_{F}^2   &  0 &  0 \\ 
0   & m_{F}^2  & 0  \\ 
0   &  0  & m_{F}^2   \\ 
\end{array} \right ),
\qquad
(m_{T}^2)_{ij} 
 = 
\left(
  \begin{array}{ccc}
m_{T(1)}^2   &  0 &  0 \\ 
0   &  m_{T(1)}^2 & 0  \\ 
0   & 0   & m_{T(2)}^2   \\ 
\end{array} \right ).
\label{eq:soft-mass-1}
\end{eqnarray}
That is, three right-handed down-type squark and 
left-handed slepton masses are degenerate, and 
first two generations of other sectors are degenerate.
These predictions  would be obvious because 
the three generations of the $F\subset(d^c,L)$ fields form  a triplet of  $S_4$, 
and the $T\subset(Q,u^c,e^c)$ fields form  a doublet and a singlet of $S_4$.
These  predictions hold exactly before $S_4\times Z_4\times U(1)_{FN}$
is broken, 
but its breaking gives next-to-leading terms in the scalar mass matrices.

Next, we study effects due to  $S_4 \times Z_4\times U(1)_{FN}$  breaking 
by $\chi_i$.
That is, we estimate corrections to the K\"ahler potential 
including  $\chi_i$. 
Since $(T_1,T_2)$ are assigned to ${\bf 2}$ and 
its conjugate representation is itself ${\bf 2}$. Similarly, 
$(F_1,F_2,F_3)$ are assigned to ${\bf 3}$ and 
its conjugation is ${\bf 3}$.  Therefore,
for the $F_{1,2,3}$ fields,  higher dimensional terms are given as
\begin{align}
\Delta K_F
&= \sum _{i=1,3} Z_{\Delta _{a_i}}^{(F)}(\Phi )
 (F_1,F_2 ,F_3 ) \otimes (F_1^c,F_2 ^c,F_3 ^c)
\otimes (\chi _i,\chi _{i+1})\otimes (\chi _i^c,\chi _{i+1}^c)/\Lambda ^2
\nonumber \\
&\ +\sum _{i=5,8,11} Z_{\Delta _{b_i}}^{(F)}(\Phi )
 (F_1,F_2 ,F_3 ) \otimes (F_1^c,F_2 ^c,F_3 ^c)
\otimes (\chi _i,\chi _{i+1},\chi _{i+2})\otimes (\chi _i^c,\chi _{i+1}^c,\chi _{i+2}^c)/\Lambda ^2
\nonumber \\
&\ + Z_{\Delta _c}^{(F)}(\Phi )
 (F_1,F_2 ,F_3 ) \otimes (F_1^c,F_2 ^c,F_3 ^c)
\otimes \chi _{14}\otimes \chi _{14}^c/\Lambda ^2
\nonumber \\
&\ +Z_{\Delta _d}^{(F)}(\Phi )
 (F_1,F_2 ,F_3 ) \otimes (F_1^c,F_2 ^c,F_3 ^c)
\otimes \Theta \otimes \Theta ^c/\bar \Lambda ^2.
\end{align}
For example, 
higher dimensional terms 
including $(\chi _1,\chi_2)$ and 
 $(\chi _5, \chi _6 ,\chi_7)$ are explicitly written  as
\begin{align}
\Delta K_F^{\left [\chi _1,\chi _5\right ]} &
= Z_{\Delta _{a_1}}^{(F)}(\Phi )
\left [\frac{\sqrt 2|\chi _1|^2}{\Lambda ^2}(|F_2 |^2-|F_3 |^2)\right ] \nonumber \\
&\ +Z_{\Delta _{b_5}}^{(F)}(\Phi )\left [\frac{2|\chi _5|^2}{\Lambda ^2}
(F_2 F_3 ^\ast +F_3 F_2^\ast +F_1F_3 ^\ast +F_3 F_1^\ast +F_1F_2^\ast 
+F_2 F_1^\ast )\right ].
\end{align}

When we take into account   corrections from all $\chi_i \chi_j^*$ 
to the K\"ahler potential, 
the soft scalar masses squared for the $F_{1,2,3}$ fields 
have the following corrections, 
\begin{equation}
(m_{F}^2)_{ij}=\begin{pmatrix}
                              m_F^2+\tilde a_{F1} ^2m_{3/2}^2 & k_Fa_5 ^2m_{3/2}^2 & k_Fa_5 ^2m_{3/2}^2 \\
                              k_Fa_5 ^2m_{3/2}^2 & m_F^2+\tilde a_{F2} ^2m_{3/2}^2 & k_Fa_5 ^2m_{3/2}^2 \\
                              k_Fa_5 ^2m_{3/2}^2 & k_Fa_5 ^2m_{3/2}^2 & m_F^2+\tilde a_{F3} ^2m_{3/2}^2
                           \end{pmatrix},
\end{equation}
where $k_F$ is a parameter of order one,
and $\tilde a_{Fk}^2(k=1,2,3)$ are linear combinations of $a_i a_j$'s.

For the $T_{1,2,3}$ fields,  
higher dimensional terms are given as
\begin{align}
\Delta K_T 
&= \sum _{i=1,3} Z_{\Delta _{a_i}}^{(T)}(\Phi )
 (T_1,T_2 ) \otimes (T_1^c,T_2 ^c)
\otimes (\chi _i,\chi _{i+1})\otimes (\chi _i^c,\chi _{i+1}^c)/\Lambda ^2
\nonumber \\
&\ +\sum _{i=5,8,11} Z_{\Delta _{b_i}}^{(T)}(\Phi )
 (T_1,T_2 ) \otimes (T_1^c,T_2 ^c) 
\otimes (\chi _i,\chi _{i+1},\chi _{i+2})\otimes (\chi _i^c,\chi _{i+1}^c,\chi _{i+2}^c)/\Lambda ^2
\nonumber \\
&\ + Z_{\Delta _c}^{(T)}(\Phi )
 (T_1,T_2 ) \otimes (T_1^c,T_2 ^c) 
\otimes \chi _{14}\otimes \chi _{14}^c/\Lambda ^2
\nonumber \\
&\ + Z_{\Delta _d}^{(T)}(\Phi )
 (T_1,T_2 ) \otimes T_3 ^c 
\otimes (\chi _1,\chi _2)/\Lambda ^2
+ Z_{\Delta _e}^{(T)}(\Phi )
 (T_1^c,T_2 ^c ) \otimes T_3 
\otimes (\chi _1^c,\chi _2^c)/\Lambda ^2
\nonumber \\
&\ +\sum_{i=1,3} Z_{\Delta _{f_i}}^{(T)}(\Phi )
 T_3 \otimes T_3 ^c
\otimes (\chi _i,\chi _{i+1})\otimes (\chi _i^c,\chi _{i+1}^c)/\Lambda ^2
\nonumber \\
&\ +\sum _{i=5,8,11} Z_{\Delta _{g_i}}^{(T)}(\Phi )
 T_3  \otimes T_3 ^c
\otimes (\chi _i,\chi _{i+1},\chi _{i+2})\otimes (\chi _i^c,\chi _{i+1}^c,\chi _{i+2}^c)/\Lambda ^2
\nonumber \\
&\ + Z_{\Delta _h}^{(T)}(\Phi )
 T_3  \otimes T_3 ^c
\otimes \chi_{14}\otimes \chi_{14}^c/\Lambda ^2
+ Z_{\Delta _i}^{(T)}(\Phi )
 (T_1,T_2 ) \otimes (T_1^c,T_2 ^c) 
\otimes \Theta \otimes \Theta ^c/\bar \Lambda ^2
\nonumber \\
&\ + Z_{\Delta _j}^{(T)}(\Phi )
 T_3  \otimes T_3 ^c
\otimes \Theta \otimes \Theta ^c/\bar \Lambda ^2.
\end{align}

In the same way, the $T_{1,2,3}$ scalar mass matrix 
can be written as
\begin{equation}
(m_{T}^2)_{ij}=\begin{pmatrix}
                              m_{T(1)}^2+\tilde a_{T11} ^2m_{3/2}^2 & \tilde a^2_{T12}m_{3/2}^2 & k_Ta_1m_{3/2}^2 \\
                              \tilde a^2_{T12}m_{3/2}^2 & m_{T(1)}^2+\tilde a_{T22} ^2m_{3/2}^2 & k_Ta_1m_{3/2}^2 \\
                              k_T^*a_1m_{3/2}^2 & k_T^*a_1m_{3/2}^2 & m_{T(2)}^2+\tilde a^2_{T33}m_{3/2}^2
                           \end{pmatrix},
\label{RR}
\end{equation}
where $k_T$ is a complex parameter whose magnitude is of order one, 
and it is the only new source of the CP violation in our model. 
The parameters $\tilde a_{Tij}^2$ are linear combinations of $a_k a_{\ell}$'s. 
In numerical analysis, we use the parameter $\Delta a_L$ 
which is given by $\Delta a_L=m_{T(2)}^2/m_{T(1)}^2-1$.


 In order to estimate the magnitude of FCNC phenomena, we move to the super-CKM basis
by diagonalizing quark and lepton mass matrices including  next-to-leading 
terms.
 For the left-handed down-type squark and slepton, we get
\begin{eqnarray}
({\tilde m}_{d_{LL}}^2)_{ij}^{(SCKM)}=U^\dagger_d (m_{ T}^2)_{ij} U_d ,
\quad
({\tilde m}_{ \ell_{LL}}^2)_{ij}^{(SCKM)}=U^\dagger_E (m_{F}^2)_{ij} U_E ,
\end{eqnarray}
and for the right-handed down-type squark and slepton as
\begin{eqnarray}
({\tilde m}_{ d_{RR}}^2)_{ij}^{(SCKM)}=V^\dagger_d (m_{ F}^2)_{ij} V_d ,
\quad
({\tilde m}_{ e_{RR}}^2)_{ij}^{(SCKM)}=V^\dagger_E (m_{ T}^2)_{ij} V_E ,
\end{eqnarray}
where the  mixing matrices $V_E$ and $U_E$ are given in  Eq. (\ref{VEUE})
in Appendix C.


Let us study scalar trilinear couplings, i.e. 
the so called A-terms.
The A-terms among left-handed and right-handed squarks (sleptons) 
and Higgs scalar fields are obtained in the gravity mediation 
as~\cite{Kaplunovsky:1993rd}
\begin{equation}
h_{IJ} {L}_J {R}_I H_K =  \sum_{K={\bar 5},\ 45}
h^{(Y)}_{IJK}{L}_J {R}_I H_K  + h^{(K)}_{IJK}{L}_J {R}_I H_K ,
\label{eq:A-term}
\end{equation}
where 
\begin{eqnarray}
h^{(Y)}_{IJK} &=& F^{\Phi_k} \langle \partial_{\Phi_k} \tilde{y}_{IJK}
\rangle ,  
\nonumber \\
h^{(K)}_{IJK}{L}_J {R}_I H_K &=& - 
\langle \tilde{y}_{LJK} \rangle {L}_J {R}_I H_K F^{\Phi_k} K^{L\bar{L}}
\partial_{\Phi_k} K_{\bar{L}I}  \\
& &  -
\langle \tilde{y}_{IMK} \rangle {L}_J {R}_I H_d F^{\Phi_k} K^{M\bar{M}} 
\partial_{\Phi_k} K_{\bar{M}J}  \nonumber  \\ 
& & -   \langle \tilde{y}_{IJK} \rangle {L}_J {R}_I H_K F^{\Phi_k} K^{H_d}
\partial_{\Phi_k} K_{H_K}, \nonumber  
\label{eq:A-term-2}
\end{eqnarray}
and $K_{H_K}$ denotes the K\"ahler metric of $H_K$.
In addition,  effective Yukawa couplings of the down-type quark
 $\tilde{y}_{IJK}$ are written as
\begin{eqnarray}
\tilde{y}_{IJK}
 = y_1
\begin{pmatrix}0 & a_9/\sqrt 2 & 0 \\ 
           0 & a_9/\sqrt 6 & 0 \\
                 0  & 0 & 0   \\
 \end{pmatrix}_{LR} 
+y_2
\begin{pmatrix} 0 & 0 & 0 \\ 
               0 & 0 & 0 \\
                 0 & 0 & a_{13} \\
 \end{pmatrix}_{LR}, 
\label{ME}
 \end{eqnarray}
then we have
\begin{eqnarray}
h^{(Y)}_{IJK}
 = \frac{y_1}{\Lambda }
\begin{pmatrix}0 & \tilde F^{a _9}/\sqrt 2 & 0 \\ 
           0 & \tilde F^{a _9}/\sqrt 6 & 0  \\
                 0  & 0 & 0   \\
 \end{pmatrix}_{LR}
+\frac{y_2}{\Lambda }
\begin{pmatrix} 0 & 0 & 0 \\ 
               0 & 0 & 0 \\
                0 & 0 & \tilde F^{a _{13}} \\
 \end{pmatrix}_{LR},
\label{ME}
\end{eqnarray}
where $\tilde F^{a_i}=F^{a_i}/a_i$ and 
$\tilde F^{a_i}/\Lambda = {\cal O}(m_{3/2})$.

 By use of lowest level of the  K\"ahler potential,
 we estimate $h^{(K)}_{IJK}$ as
\begin{equation}
h^{(K)}_{IJK} = \tilde y_{IJK} (A^R_I+A^L_J),
\end{equation}
where we assume $A^L_1=A^L_2=A^L_3 
=F^{\tilde a_i}/(a_i\Lambda) \simeq \mathcal{O}(m_{3/2})$. 
The magnitudes of $A^R_1= A^R_2$ and $A^R_3$ are also $\mathcal{O}(m_{3/2})$.
Furthermore, we should take into account next-to-leading terms of 
the K\"ahler potential including $\chi_i$.
These correction terms appear all entries so that their  magnitudes
 are suppressed in ${\cal O}(\tilde a)$
compared with the leading term.  Then, we obtain 
\begin{equation}
 (m_{d_{LR}}^2)_{ij} \simeq 
(m_{\ell_{LR}}^2)_{ij}^\dagger \simeq
m_{3/2}
\begin{pmatrix}
\tilde a_{LR11} ^2v_d & c_1\frac{\sqrt{3}m_{s(\mu)}}{2} & \tilde a_{LR13} ^2v_d \\ 
\tilde a_{LR21} ^2v_d & c_1\frac{m_{s(\mu)}}{2} & \tilde a_{LR23} ^2v_d \\
\tilde a_{LR31} ^2v_d & \tilde a_{LR32} ^2v_d & c_2m_{b(\tau)} 
\end{pmatrix}_{LR} ,
\end{equation}
where  $\tilde a_{LRij}^2$ are linear combinations of $a_k a_{\ell}$'s,
and $c_1$ and $c_2$ are of order one parameters.
Moving to the super-CKM basis, we have
\begin{align}
 ({\tilde m}_{{ d}_{LR}}^2)^{(SCKM)}_{ij} =U_d^\dagger  (m_{d_{LR}}^2)_{ij} V_d
 \simeq m_{3/2}
\begin{pmatrix}
\mathcal{O} \left (\tilde a^2v_d\right ) 
& \mathcal{O} \left (\tilde a^2v_d\right ) 
& \mathcal{O} \left (\tilde a^2v_d\right ) \\ 
\mathcal{O} \left (\tilde a^2v_d\right ) 
& \mathcal{O}(m_s) & \mathcal{O} \left(\tilde a ^2v_d\right ) \\
\mathcal{O} \left (\tilde a^2v_d\right ) 
& \mathcal{O} \left (\tilde a^2v_d\right ) & \mathcal{O}  (m_b)
\end{pmatrix}.
\end{align}
Similarly, for the charged lepton, 
\begin{align}
 ({\tilde m}_{{ \ell}_{LR}}^2)^{(SCKM)}_{ij} =U_E^\dagger  (m_{\ell_{LR}}^2)_{ij} V_E
 \simeq m_{3/2}
\begin{pmatrix}
\mathcal{O} \left (\tilde a^2v_d\right ) 
& \mathcal{O} \left (\tilde a^2v_d\right ) 
& \mathcal{O} \left (\tilde a^2v_d\right ) \\ 
\mathcal{O} \left (\tilde a^2v_d\right ) 
& \mathcal{O}(m_\mu) & \mathcal{O} \left(\tilde a ^2v_d\right ) \\
\mathcal{O} \left (\tilde a^2v_d\right ) 
& \mathcal{O} \left (\tilde a^2v_d\right ) & \mathcal{O}  (m_\tau)
\end{pmatrix}.
\end{align}


\subsection{Renormalization group effect}

In the framework of the supergravity, 
soft masses for all scalar particles have the common scale denoted 
by $m_{\text{SUSY}}$, and gauginos also have the common scale $m_{1/2}$. 
Therefore, at the GUT scale $m_\text{GUT}$, we take 
\begin{eqnarray}
M_1 (m_\text{GUT}) = M_2 (m_\text{GUT}) = M_3 (m_\text{GUT}) = m_{1/2} \; .
\end{eqnarray}
Effects of the renormalization group running lead at the scale $m_W$ 
to following masses for gauginos,
\begin{eqnarray}
\label{gaugino}
M_i(m_W)\simeq\dfrac{\alpha _i(m_W)}{\alpha_i (m_\text{GUT})}M_i(m_\text{GUT}).
\end{eqnarray}
Taking into account the renormalization group effect \cite{Martin:1993zk}
on the average mass scale in $m_{e_{L}}^2$, $m_{e_{R}}^2$, 
$m_{q_{L}}^2$, and $m_{d_{R}}^2$ with neglecting Yukawa couplings, we have
\begin{eqnarray}
\begin{split}
\label{smass}
m_{e_{L}}^2(m_W)
&\simeq m_L^2(m_\text{GUT})+0.5M_2^2(m_\text{GUT})
+0.04M_1^2(m_\text{GUT}) \simeq m_{\text{SUSY}}^2 +0.54 m_{1/2}^2 ,
\\
m_{e_{R}}^2(m_W)
&\simeq m_R^2(m_\text{GUT})+0.15M_1^2(m_\text{GUT}) 
\simeq m_{\text{SUSY}}^2 +0.15 m_{1/2}^2 \; ,
\\
m_{q_L}^2(m_W)
&\simeq m_R^2(m_\text{GUT})+0.004M_1^2(m_\text{GUT})
+0.4 M_2^2(m_\text{GUT})+3.6 M_3^2(m_\text{GUT})
\\
&\simeq m_{\text{SUSY}}^2 +4.1 m_{1/2}^2 \; ,
\\
m_{d_R}^2(m_W)
&\simeq m_R^2(m_\text{GUT})+0.015M_1^2(m_\text{GUT})
+3.6 M_3^2(m_\text{GUT})
\simeq m_{\text{SUSY}}^2 +3.7 m_{1/2}^2 \; .
\end{split}
\end{eqnarray}

For Yukawa couplings, the $b-\tau$ unification is realized 
at the leading order in our model, 
however, the $b-\tau$ unification is deviated when we include
 the next-to-leading order mass operators  due to terms including $H_{45}$, 
see Ref. \cite{Ishimori:2010su} for the detail. 
In that paper, we have calculated the renormalization group 
equations and observed fermion masses at the weak scale 
can be obtained when $\tan\beta$ is larger than two. 
Hereafter, we take $\tan\beta=3$ on the numerical analysis.


\section{Numerical Analysis}
\label{sec:analysis}

In this section, we perform numerical analysis to show that 
the $S_4$ flavor model presented in the previous section can explain 
the like-sign charge asymmetry in the $B_s-\bar B_s$ system.  
In order for this calculation, we first define the MI parameters for down-type squarks 
$\delta_d^{LL}$, $\delta_d^{LR}$, $\delta_d^{RL}$, and $\delta_d^{RR}$
and for sleptons $\delta_\ell^{LL}$, $\delta_\ell^{LR}$,
 $\delta_\ell^{RL}$ and, $\delta_e^{RR}$ as 
\begin{eqnarray}
\begin{split}
&m_{\tilde q}^2
 \begin{pmatrix} \delta_d^{LL} & \delta_d^{LR} \\ 
                   \delta_d^{RL}    & \delta_d^{RR}  \\
 \end{pmatrix}
=
\begin{pmatrix} ({\tilde m}_{{ d}_{LL}}^2)^{(SCKM)} & ({\tilde m}_{{ d}_{LR}}^2)^{(SCKM)} \\ 
                (   {\tilde m}_{{ d}_{RL}}^2  )^{(SCKM)}  &( {\tilde m}_{{ d}_{RR}}^2)^{(SCKM)}  \\
 \end{pmatrix}
- \text{diag}(m_{\tilde q}^2) \ ,
\\
&m_{\tilde \ell}^2
 \begin{pmatrix} \delta_\ell^{LL} & \delta_\ell^{LR} \\ 
                   \delta_\ell^{RL}    & \delta_e^{RR}  \\
 \end{pmatrix}
=
\begin{pmatrix} ({\tilde m}_{{ \ell}_{LL}}^2)^{(SCKM)} & ({\tilde m}_{{ \ell}_{LR}}^2)^{(SCKM)} \\ 
               (    {\tilde m}_{{ \ell}_{RL}}^2)^{(SCKM)}    & ({\tilde m}_{{ e}_{RR}}^2)^{(SCKM)}  \\
 \end{pmatrix}
- \text{diag}(m_{\tilde \ell}^2) \ ,
\end{split}
\end{eqnarray}
where $m_{\tilde q}$ 
and $m_{\tilde\ell}$ are average squark and slepton masses with the values 
given below.

In the numerical analysis, we fix the following parameters
\be
m_{3/2}&=&430~{\rm GeV},~m_{\tilde q}=880~{\rm GeV},~m_{\tilde \ell}=520~{\rm GeV},\nonumber \\
M_1&=&135~{\rm GeV},~M_2=270~{\rm GeV},~M_3\equiv m_{\tilde g}=1~{\rm TeV}, 
\ee 
which are derived from the universal relation at the GUT scale
 \be
m_{1/2}(m_\text{GUT})
=m_{3/2}(m_\text{GUT})
=m_{\rm SUSY}(m_\text{GUT})=&430~{\rm GeV},
\ee 
by through the renormalization group effect discussed in the section III-C. 
This universal value is taken to be consistent with 
the lower bound of the gluino mass, which has been reported recently
at Atlas Collaboration of LHC
\cite{Aad:2011hh,daCosta:2011qk,Collaboration:2011ks}.
For the other parameters, we assume the following regions: 
\be
\mu&=&[500,1000]~{\rm GeV},~\Delta a_L=[-0.5,5],~
|k_T a_1|=[0,2],~{\rm arg}(k_T a_1)=[-\pi,\pi],\nonumber \\
a_5&=&[0,0.001],~\tilde a_{T12}=[0,0.1],~\tilde a_{LRij}=[0,0.01], 
\label{para3}
\ee
with $\tan \beta=3,~c_{1,2}=1,$ and $k_F=1$. 
In Eq.(\ref{para3}), the number of left-handed  and right-handed sides in  braces
denote the minimal and maximal values, respectively.
In our calculation, we neglect the diagonal elements of scalar masses $\tilde a_{F(1,2,3)}$ and $\tilde a_{T(11,22,33)}$. 
The leading contribution to the parameters $\tilde a$ in the soft-terms are 
$a_1$ 
 as $\tilde a_{LRij} \simeq \sqrt{a_1 a_5}$ and $\tilde a_{T12}\simeq a_1$. 
As given in Appendix \ref{sec:formulae}, the SUSY contribution to $M_{12}^{s,SUSY}$ is estimated as 
\be
M_{12}^{s,SUSY}\simeq -\frac{\alpha_S^2}{216 m_{\tilde q}^2}\frac{2}{3}M_{B_s}f_{B_s}^2 
\Bigl\{-0.59 \left[ (\delta_d^{LL})^2_{23}+(\delta_d^{RR})^2_{23}\right]
+31  (\delta_d^{LL})_{23} (\delta_d^{RR})_{23} \nonumber \\
-9.4 \left[ (\delta_d^{LR})^2_{23}+(\delta_d^{RL})^2_{23}\right]
+7.9 (\delta_d^{LR})_{23} (\delta_d^{RL})_{23}\Bigr\}, 
\ee
for $x=m_{\tilde g}^2/m_{\tilde q}^2\simeq 1.3$, and similar for $B_d$ mixing. 
The coefficients in front of MI parameters for $K$ meson mixing are 
$-0.59,554,-183, 114$, respectively. 
Since the $LR$ terms $(\delta^{LR,RL}_d)_{ij}$ are strictly constrained by $b \to s\gamma$ as seen in  Appendix \ref{sec:formulae}, 
the $(\delta_d^{LL(RR)})_{ij}$ terms  
gives larger contribution to $M_{12}^{s,SUSY}$. Among them, 
since $(\delta_d^{RR})_{ij}\simeq k_F a_5^2\lsim 10^{-6}$ in our parameter region given in Eq.(\ref{para3}), 
the first term $(\delta_d^{LL})_{ij}^2$ gives the dominant contributions.  
The approximation form of the $LL$ parameters $(\delta_d^{LL})_{ij}$ are given by  
\be
(\delta_d^{LL})_{12} &\simeq& \theta_{13}^d \theta_{23}^d  \Delta a_L
-\frac{m_{3/2}^2}{m_{\tilde q}^2} \left( \theta_{13}^d \frac{1+\sqrt{3}}{2}k_T^* a_1+\theta_{23}^d\frac{1-\sqrt{3}}{2}k_T a_1\right)
\nonumber \\
&\simeq&-\frac{m_{3/2}^2}{m_{\tilde q}^2} \sqrt{2}\left(\theta^d_{13}V_{ud}k_T^* a_1-\theta_{23}^d V_{us}k_T a_1 \right), 
\label{d12aprx}\\
(\delta_d^{LL})_{13} &\simeq& -\theta_{13}^d \Delta a_L+\frac{m_{3/2}^2}{m_{\tilde q}^2}
\left(\frac{1-\sqrt{3}}{2}-\theta_{12}^d \frac{1+\sqrt{3}}{2}\right)k_T a_1
\simeq -\frac{m_{3/2}^2}{m_{\tilde q}^2} \sqrt{2}V_{us} k_T a_1,
\label{d13aprx}\\
(\delta_d^{LL})_{23} &\simeq& -\theta_{23}^d \Delta a_L+\frac{m_{3/2}^2}{m_{\tilde q}^2}
\left(\frac{1+\sqrt{3}}{2}+\theta_{12}^d \frac{1-\sqrt{3}}{2}\right)k_T a_1
\simeq \frac{m_{3/2}^2}{m_{\tilde q}^2} \sqrt{2}V_{ud}k_T a_1, 
\label{d23aprx}
\ee
where in the last approximation of each expression, we have neglected the first term 
proportional to $\theta_{13,23}^d\simeq 0.005$ and $\Delta a_L\sim 1$. 
Notice that the MI parameters are expressed in terms of the CKM elements, and 
that both $(\delta_d^{LL})_{13}$ and $(\delta_d^{LL})_{23}$ have the 
same phase structure $k_T a_1$, which is only the new source of the CP violation in our model. 
These are the typical  feature of our  $S_4$ flavor model. 
By using these expressions, one finds that the $K^0-\bar K^0$ mixing induced by $(\delta_d^{LL})_{12}$ 
is more suppressed by additional factor $\theta_{ij}^d$. 

The cEDM for the strange quark is estimated from the formula in Appendix  \ref{sec:formulae} as 
\be
e d^C_s \sim 10^{-20}\text{Im}[(\delta_d^{LL})_{23}(\delta _d^{LR})_{33}(\delta _d^{RR})_{32}]~e{\rm cm}
\sim 10^{-28}{\rm Im}[(\delta _d^{LL})_{23}]~e{\rm cm}, 
\ee
for $x\simeq 1.3$, $(\delta_{33}^d)_{LR}\sim 10^{-2}$ and $(\delta_{32}^d)_{RR}\sim10^{-6}$. Therefore, 
$(\delta _{23}^d)_{LL}$ is not constrained by cEDM. 
As for the $b \to s \gamma$ process, 
one can see that $(\delta^d_{23})_{LR}$ should be strongly suppressed while 
$(\delta^d_{23})_{LL,RR}$ have an additional suppression factor $m_b/m_{\tilde g}\sim 10^{-3}$. 
In our numerical calculation, we take $\tilde a_{LRij}\lsim 0.01$ 
so that $b \to s \gamma$ is well suppressed, and the allowed region of 
$|(\delta^{LL}_d)_{23}|$ is also small enough as mentioned below.

First we discuss the allowed regions of the parameters $(h_s,h_d),~(h_s,\sigma_s),~(h_d,\sigma_d)$ 
defined in Eq.(\ref{hsigmaDelta}). 
In our model, the parameters $h_{d,s}e^{2 i \sigma_{d,s}}$ are estimated as  
\be
h_d e^{2 i \sigma_d}&=&\frac{M_{12}^{d,SUSY}}{M_{12}^{d,SM}}\simeq 
(27-i 25)( \delta_d^{LL})_{13}^2,~
\label{ddelta}\\
h_s e^{2 i \sigma_s}&=&\frac{M_{12}^{s,SUSY}}{M_{12}^{s,SM}}\simeq 
(1.7+i 0.06)( \delta_d^{LL})_{23}^2, 
\label{sdelta}
\ee
where the MI parameters $(\delta_d^{LL})_{ij}$ reflect the flavor symmetry, 
while the factors $(27-i 25)$ and $(1.7+i 0.06)$ do not. 
The ratio of $(27-i 25)/(1.7 +i 0.06)$ is related to the CKM elements as 
$(27-i 25)/(1.7 +i 0.06)\simeq M_{12}^{s,SM}/M_{12}^{d,SM}\simeq (V_{ts}^*/V_{td}^*)^2$. 
We obtain the ratio of $h_d$ and $h_s$ as 
\be
\frac{h_d}{h_s}\simeq \frac{|27-i 25|}{|1.7+i 0.06|}\frac{|V_{us}|^2}{|V_{ud}|^2}
\simeq \frac{|V_{ts}|^2}{|V_{td}|^2}\frac{|V_{us}|^2}{|V_{ud}|^2}\simeq 1.
\label{hdoverhs}
\ee
Therefore, the fact that the region $h_d \simeq h_s$ is favored 
reflects the flavor structure of the $S_4$ flavor model. 
The CP violation phase $\phi_s$ is given by 
\be
\phi_s \simeq{\rm arg}\left[ -(1+h_s e^{2 i \sigma_s})\right],
\ee
with neglecting the SM contribution. 
The CP phase $\sin \phi_s$ is bounded as $|\sin \phi_s|\lsim h_s$ for $h_s<1$, and has 
the negatively-maximal value $\sin \phi_s \simeq -h_s$ 
at $\sigma_s \simeq 120^\circ$. 
This corresponds to the best-fit value $(h_s,\sigma_s)=(0.5,120^\circ)$ of 
Eq.(\ref{bestfit})\cite{Ligeti:2010ia}. 

Fig.1 shows the plot in the $\phi_s-A_{sl}^b$ plane. 
The horizontal and vertical lines are the 
experimental values of one-dimensional likelihood analysis \cite{CDF2010}
\be
\phi_s=[-1.8,0.4]~({\rm rad}),~{\rm at}~95\%~{\rm C.L.}, 
\ee
and 2$\sigma$ range of $A_{sl}^b$ in Eq.(\ref{Abexp}), respectively. 
The blue (dark gray) and orange (light gray) regions denote 2$\sigma$ and 3$\sigma$ regions of 
$A_{sl}^b$ in Eq.(\ref{Abexp}), respectively. 
By using Eq.(\ref{aslq}) and $|1+h_s {\rm exp}({2 i\sigma_s})|\sim (1+0.5 {\rm exp}[2 i 120^\circ])\simeq 0.8$, 
we obtain $-a_{sl}^s\gsim 10^{-3}$, and 
similar for $a_{sl}^d$. As a consequence we obtain the like-sign charge asymmetry 
as $-A_{sl}^b\gsim 10^{-3}$, 
which is within 2$\sigma$ range of $A_{sl}^b$ of Eq.(\ref{Abexp}).  

Fig.2 shows the allowed region in the ${\rm Re}(\delta_d^{LL})_{23}-{\rm Im}(\delta_d^{LL})_{23}$ plane. 
One finds from Eq.(\ref{sdelta}) that in order to obtain the best fit value 
$(h_s,\sigma_s)=(0.5,120^\circ)$, the sign of ${\rm Re}(\delta_d^{LL})_{23}$ and ${\rm Im}(\delta_d^{LL})_{23}$ 
must be opposite from each other, with 
${\rm Re}(\delta_d^{LL})_{23} \simeq \pm 0.3$ and 
${\rm Im}(\delta_d^{LL})_{23} \simeq \mp 0.4$. 
This allowed region $|(\delta_d^{LL})_{23}| \lsim 0.5$ is small enough to suppress $b \to s \gamma$. 
The similar figure is drawn in the ${\rm Re}(\delta_d^{LL})_{13}-{\rm Im}(\delta_d^{LL})_{13}$ plane 
with $|V_{td}/V_{ts}| \simeq 0.22$ times smaller area. 


\begin{figure}[t]
{\includegraphics[width=7cm]{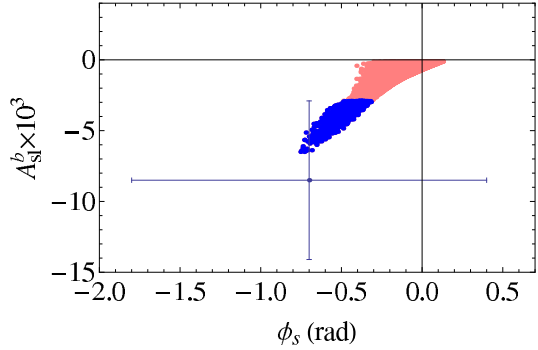}}
\caption{Allowed region in the $\phi_s-A_{sl}^b$ plane. The blue (dark gray) and orange (light gray) regions denote 2$\sigma$ and 3$\sigma$ regions of 
$A_{sl}^b$ in Eq.~(\ref{Abexp}). The blue (black) error bars of the horizontal and vertical lines are experimental values of
 $2\sigma $ region of $\phi _s$ and $A_{sl}^b$.}
\end{figure}

\begin{figure}[t]
{\includegraphics[width=7cm]{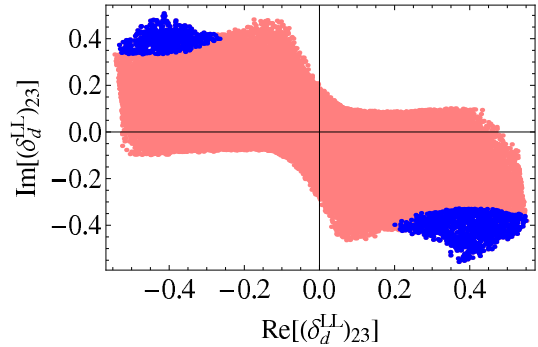}}
\caption{Allowed region in the ${\rm Re}(\delta_d^{LL})_{23}-{\rm Im}(\delta_d^{LL})_{23}$ plane. 
The blue (dark gray) and orange (light gray) regions denote 2$\sigma$ and 
3$\sigma$ regions of $A_{sl}^b$ in Eq.~(\ref{Abexp}).}
\end{figure}

\begin{figure}[t]
{\includegraphics[width=7cm]{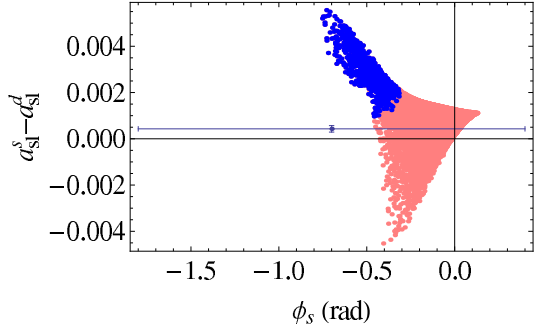}}
\caption{Allowed region in the $(a_{sl}^s-a_{sl}^d)-\phi_s$ plane. 
The blue (dark gray) and orange (light gray) regions denote 2$\sigma$ and 3$\sigma$ 
regions of $A_{sl}^b$ in Eq.~(\ref{Abexp}). The blue error bars of 
the horizontal and vertical lines are experimental values of $2\sigma $
 region of $\phi _s$ and $(a_{sl}^s-a_{sl}^d)$. 
The horizontal line is experimental values of 2$\sigma$ region for $\phi_s$, 
and the vertical line is the SM prediction of $(a_\text{sl}^s-a_\text{sl}^d)$.}
\end{figure}

\begin{figure}[t]
{\includegraphics[width=7cm]{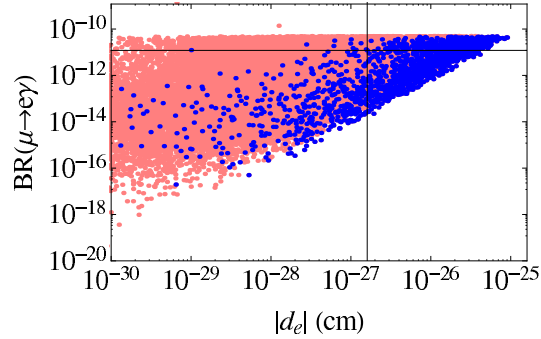}}
\caption{Allowed region in the $|d_e|-BR(\mu \to e \gamma)$ plane. 
The blue (dark gray) and orange (light gray) regions denote 2$\sigma$ and 3$\sigma$ regions of $A_{sl}^b$ in Eq.~(\ref{Abexp}). 
The horizontal and vertical lines are experimental bounds of $|d_e|$ and $BR(\mu \to e \gamma)$.}
\end{figure}

In Fig.3, we predict the difference $a_{sl}^s-a_{sl}^d$, 
which will be measured at LHCb, as a function of 
$\phi_s$. 
The SM prediction \cite{Nierste:2011ti}
$a_{sl}^{s,SM}-a_{sl}^{d,SM}=(4.3 \pm 0.7)\times 10^{-4}$
is also shown. We predict that  
$a_{sl}^s-a_{sl}^d \simeq (1-5) \times 10^{-3}$ 
in the 2$\sigma$ region of $A_{sl}^b$. This will be a good test 
for our $S_4$ flavor model. 

Since our model is based on the $SU(5)$ GUT, above contributions in  the
quark sector affect to the lepton sector. Therefore, 
 sleptons contribute to the LFV processes 
and EDM of the electron
\cite{Borzumati:1986qx,Hisano:1995nq},
in which the experimental measurements  give the upper bounds
\cite{Adam:2009ci,Regan:2002ta,DeMille}.
The Fig.4 shows the relation of ${\rm BR}(\mu\to e\gamma)$ and 
the electron EDM. 
Within the MI parameters, $ (\delta_e^{RR})_{ij}\sim(\delta_d^{LL})_{ji}$ except for 
$(i,j)=(1,2),(2,1)$ are relatively large in our model. 
Therefore one finds from Appendix E that the $(\delta_e^{RL})_{21}$ term 
in $A_L^{21}$, which is enhanced by $M_1/m_{\mu}\sim 10^3$, 
mainly contributes to $\mu \to e \gamma$ process. 
As for the electron EDM, the terms with one small MI parameters dominates. 
The largest contributions are approximately estimated as 
\be
{\rm BR}(\mu \to e \gamma)&\simeq& \frac{48 \pi^3 \alpha}{G_F^2} 
\left| \frac{\alpha_1}{4 \pi} \frac{(\delta_e^{RL})_{21}}{m_{\tilde \ell}^2}\left( \frac{M_1}{m_{\mu}}\right)
2 f_{2n}(x_1)\right|^2 \nonumber \\
&\simeq& 3 \times 10^{-11} \left( \frac{520{\rm GeV}}{m_{\tilde \ell}}\right)^4 
\left( \frac{M_1}{135{\rm GeV}}\right)^2 \left(\frac{|(\delta_e^{RL})_{21}|}{10^{-5}}\right)^2,\\
|d_e/e| &\simeq& \frac{\alpha_1}{4 \pi}\frac{M_1}{m_{\tilde \ell}^2}\Bigl|{\rm Im}
[(\delta_{\ell}^{LR})_{13}(\delta_e^{RR})_{31}]f_{3n}(x_1)\nonumber \\
&+&{\rm Im}[(\delta_{\ell}^{LR})_{12}(\delta_e^{RR})_{23}(\delta_e^{RR})_{31}
+(\delta_{\ell}^{LR})_{13}(\delta_e^{RR})_{33}(\delta_e^{RR})_{31}]f_{4n}(x_1)\Bigr| \nonumber \\
&\simeq& 1\times 10^{-26}cm  \times 
\left( \frac{M_1}{135{\rm GeV}}\right)\left( \frac{520{\rm GeV}}{m_{\tilde \ell}}\right)^2 
\Bigl[
\left( \frac{(\delta_{\ell}^{LR})_{13}}{10^{-5}}\right)\left( \frac{(\delta_e^{RR})_{31}}{0.1}\right)
+\cdots \Bigr].
\ee
The value of $|(\delta_e^{RL})_{ij}|$ is of order 
$m_{3/2}v_d/m_{\tilde \ell}^2\times \tilde a_{LRij}^2\lsim 10^{-5}$.  
Therefore we find that ${\rm BR}(\mu \to e \gamma)$ and electron EDM 
can be close to the present experimental bound as shown in the figure. 

The $b-s$ transition by $(\delta_d^{LL})_{23}$ in the quark sector simultaneously 
induce the LFV $\tau$ decay $\tau \to \mu \gamma$ by $(\delta_e^{RR})_{32}$.  
The dominant contribution is estimated from Appendix E as 
\be
{\rm BR}(\tau \to \mu \gamma)&\simeq& {\rm BR}(\tau \to \mu \nu_{\tau} \bar \nu_{\mu})
\frac{48 \pi^3 \alpha}{G_F^2}
\left( \frac{\alpha_1}{4 \pi}\right)^2
\left[ \frac{(\delta_e^{RR})_{32}}{m_{\tilde \ell}^2}\mu M_1 \tan \beta
\left( \frac{f_{3n}(x_1)}{m_{\tilde \ell}^2}-\frac{2f_{2n}(x_1,x_{\mu})}{\mu^2-M_1^2}\right)\right]^2 
\nonumber \\
&\simeq& 10^{-8}\left( \frac{520{\rm GeV}}{m_{\tilde \ell}}\right)^8 \left( \frac{M_1}{135{\rm GeV}}\right)^2
 \left( \frac{\tan \beta}{3}\right)^2\left( \frac{\mu}{500{\rm GeV}}\right)^2
| (\delta_e^{RR})_{32}|^2, 
\ee
and similar for $\tau \to e \gamma$ decay. 
Therefore for large $\mu$ term, $\tau \to \mu \gamma$ can be close to 
present upper bound given in Eq.(\ref{lfvbound}).  
By using the expression Eqs.(\ref{d13aprx}) and (\ref{d23aprx}), 
we obtain the relation of ${\rm BR}(\tau \to \mu \gamma)$ and ${\rm BR}(\tau \to e \gamma)$ 
depending on the Cabibbo angle $\lambda_c\simeq 0.22$ as follows: 
\be
\frac{{\rm BR}(\tau \to e \gamma)}{{\rm BR}(\tau \to \mu\gamma)}
\simeq \frac{|(\delta_e^{RR})_{31}|^2}{|(\delta_e^{RR})_{32}|^2}
\simeq \frac{|(\delta_d^{LL})_{13}|^2} {|(\delta_d^{LL})_{23}|^2}
\simeq \frac{|V_{us}|^2}{|V_{ud}|^2}\simeq \lambda_c^2\simeq 0.05.
\ee
Therefore we conclude that there exist the parameter region which can 
explain the like-sign dimuon asymmetry $A_{sl}^b$ 
in the $S_4$ flavor model, and in this case we predict that 
the LFV $\tau \to \mu \gamma$ decay can be so large that 
future experiments will reach, and the ratio of LFV of $\tau$ decays, 
$\tau \to e \gamma$ and $\tau \to \mu \gamma$, depends on the Cabibbo angle $\lambda_c$. 



\section{Summary}

Recently the D$\O$ Collaboration reported the like-sign dimuon charge 
asymmetry 
$A_{sl}^b$ in $b \bar b\to \mu^{\pm}\mu^{\pm}X$ decay processes. 
Their result shows 
3.2$\sigma$ deviation from the standard model prediction. One promising 
interpretation of this result is that there exist additional contribution of 
new physics 
to the CP violation in $B_s-\bar B_s$ mixing process. 
In the effective Hamiltonian of the neutral $B_s$ meson system, 
there are three physical quantities $|\Gamma_{12}^s|$, 
$|M_{12}^s|$ and the CP phase $\phi_s={\rm arg}(-M_{12}^s/\Gamma_{12}^s)$. 
In order to obtain large CP asymmetry in the neutral $B_s$ meson system, 
additional contributions from new physics to at least one of these three quantities are required. 
Within these possibilities, one can consider new physics that the absorptive part $\Gamma_{12}^s$ 
can be enhanced. However in general supersymmetric models, 
the gluino-squark box diagrams  
give the dominant contributions to $B_s-\bar B_s$ mixing, which do not affect $|\Gamma_{12}^s|$.  
Therefore in those models, new physics contributes to $|M_{12}^s|$ and 
$\phi_s$. 

In this paper we have considered an $SU(5)$ SUSY GUT with 
$S_4$ flavor symmetry. 
In this model, the Cabibbo angle, $\lambda_c \sim \sin 15^\circ$, 
of the quark sector is given by 
a difference of $45^\circ$ from up sector and $60^\circ$ from down sector 
due to the Clebsch-Gordan coefficients at the leading order. 
As for the lepton sector, 
the tri-bimaximal form is generated in neutrino sector.
 These are consequences of 
the $S_4$ flavor symmetry. Since the matter multiplet 
$T(10)$ and $F(\bar 5)$ are 
embedded into ${\bf 2}+{\bf 1}$ and ${\bf 3}$ of the $S_4$ group, respectively, 
the scalar masses of 
right-handed down-type squark and left-handed slepton are degenerated at the 
leading order, while those of $T_{1,2,3}$ fields are degenerated 
in the first two generations. 
Moreover for scalar mass matrix of $T_{1,2,3}$ fields, the relation 
$(m^2_{ T})_{13}=(m^2_{T})_{23}\propto k_T a_1$ holds 
due to the $S_4$ symmetry.  
The factor $k_T a_1$ in the scalar mass matrix is assumed to be the only 
additional complex parameter in our model, 
which is responsible for the CP violation in the neutral $B_s$ meson system via gluino-squark box diagrams. 
As a consequence, the mass-insertion parameters $(\delta_d^{LL})_{13}$ and 
$(\delta_d^{LL})_{23}$ have approximately the structure of $V_{us} k_T a_1$ and $V_{ud} k_T a_1$, 
respectively.  

We have shown that the like-sign charge asymmetry $A_{sl}^b$ is in the 
2$\sigma$ range of the combined result of D$\O$ and CDF measurements. 
Since the relation between two CP phases $\sin \phi_{d}\simeq \sin \phi_s$ holds  
due to $S_4$ flavor symmetry, and it can be large, we obtain large wrong-sign and like-sign 
asymmetry: $|a_{sl}^{d,s}|\sim |A_{sl}^b| \sim 10^{-3}$. 
The SUSY contributions in the quark sector affect to the lepton sector 
because of 
the $SU(5)$ GUT relation $(\delta^{LL}_d)_{ij} \simeq (\delta^{RR}_e)_{ji}$. 
In the parameter region allowed by $A_{sl}^b$, we have two predictions 
in the leptonic processes: 
(i) Both ${\rm BR}(\mu \to e \gamma)$ and the electron EDM 
are close to the present upper bound. 
Therefore, the MEG experiment \cite{Adam:2009ci} will be 
a good test of our model. 
(ii) The LFV $\tau$ decays, $\tau \to \mu\gamma$ and $\tau \to e \gamma$, are 
related to each other via the Cabibbo angle $\lambda_c$: 
${\rm BR}(\tau \to e\gamma)/{\rm BR}(\tau\to \mu\gamma)\simeq \lambda_c^2$. 
This is also testable at future experiments such as superKEKB.

\vspace{0.5cm}

\noindent
{ \bf Acknowledgments }\\
H.I. and Y.S are supported by Grand-in-Aid for Scientific Research,
No.21.5817 and No.22.3014, respectively,
 from the Japan Society of Promotion of Science.
The work of Y.K. is supported by the ESF grant No. 8090 and 
Young Researcher Overseas Visits Program for Vitalizing Brain Circulation Japanese in JSPS.   
The work of M.T. is  supported by the
Grant-in-Aid for Science Research, No. 21340055,
from the Ministry of Education, Culture,
Sports, Science and Technology of Japan.
\def\mat2#1#2#3#4{\left(\begin{array}{cc}#1 & #2 \\#3 & #4 \\\end{array}\right)}

\def\Mat3#1#2#3#4#5#6#7#8#9{\left(\begin{array}{ccc}#1 & #2 & #3 \\#4 & #5 & #6 \\#7 & #8 & #9 \\\end{array}\right)}

\appendix{\section{Multiplication rule of $S_4$}}
The $S_4$ group has 24 distinct elements and irreducible representations 
${\bf 1},~{\bf 1}',~{\bf 2},~{\bf 3}$, and ${\bf 3}'$.
All of the $S_4$ elements are written by products of 
the generators $b_1$ and $d_4$, 
which  satisfy
\begin{equation}
(b_1)^3=(d_4)^4=e,\quad d_4(b_1)^2d_4=b_1,\quad d_4b_1d_4=b_1(d_4)^2b_1\ .
\end{equation}
These generators are represented on 
${\bf 2}$, ${\bf 3}$ and ${\bf 3}'$ as follows,
\begin{equation}
b_1=\mat2{\omega}{0}{0}{\omega^2},\quad
d_4=\mat2{0}{1}{1}{0}, \qquad {\rm~~on~~{\bf 2}},
\end{equation}
\begin{equation} 
b_1=\Mat3{0}{0}{1} {1}{0}{0} {0}{1}{0},\quad d_4=\Mat3{-1}{0}{0} 
{0}{0}{-1} {0}{1}{0}, \qquad {\rm~~on~~{\bf 3}},
\end{equation}
\begin{equation} 
b_1=\Mat3{0}{0}{1} {1}{0}{0} 
{0}{1}{0},\quad d_4=\Mat3{1}{0}{0} {0}{0}{1} {0}{-1}{0}, 
\qquad {\rm~~on~~{\bf 3}'}.
\end{equation}
The multiplication rule depends on the basis.
We present  the multiplication rule, 
which is used in this paper:
\begin{align}
\begin{pmatrix}
a_1 \\
a_2
\end{pmatrix}_{\bf 2} \otimes  \begin{pmatrix}
                                      b_1 \\
                                      b_2
                                  \end{pmatrix}_{\bf 2}
 &= (a_1b_1+a_2b_2)_{{\bf 1}}  \oplus (-a_1b_2+a_2b_1)_{{\bf 1}'} 
  \oplus  \begin{pmatrix}
             a_1b_2+a_2b_1 \\
             a_1b_1-a_2b_2
         \end{pmatrix}_{{\bf 2}\ ,} \\
\begin{pmatrix}
a_1 \\
a_2
\end{pmatrix}_{\bf 2} \otimes  \begin{pmatrix}
                                      b_1 \\
                                      b_2 \\
                                      b_3
                                  \end{pmatrix}_{{\bf 3}}
 &= \begin{pmatrix}
          a_2b_1 \\
          -\frac{1}{2}(\sqrt 3a_1b_2+a_2b_2) \\
          \frac{1}{2}(\sqrt 3a_1b_3-a_2b_3)
      \end{pmatrix}_{{\bf 3}} \oplus \begin{pmatrix}
                                        a_1b_1 \\
                                        \frac{1}{2}(\sqrt 3a_2b_2-a_1b_2) 
\\
                                        -\frac{1}{2}(\sqrt 3a_2b_3+a_1b_3)
                                   \end{pmatrix}_{{\bf 3}'\ ,} \\
\begin{pmatrix}
a_1 \\
a_2
\end{pmatrix}_{\bf 2} \otimes  \begin{pmatrix}
                                      b_1 \\
                                      b_2 \\
                                      b_3
                                  \end{pmatrix}_{{\bf 3}'}
&= \begin{pmatrix}
         a_1b_1 \\
         \frac{1}{2}(\sqrt 3a_2b_2-a_1b_2) \\
         -\frac{1}{2}(\sqrt 3a_2b_3+a_1b_3)
     \end{pmatrix}_{{\bf 3}} \oplus
      \begin{pmatrix}
                                      a_2b_1 \\
                                      -\frac{1}{2}(\sqrt 3a_1b_2+a_2b_2) \\
                                      \frac{1}{2}(\sqrt 3a_1b_3-a_2b_3)
                                  \end{pmatrix}_{{\bf 3}'\ ,} \\
\begin{pmatrix}
a_1 \\
a_2 \\
a_3
\end{pmatrix}_{{\bf 3}} \otimes  \begin{pmatrix}
                                      b_1 \\
                                      b_2 \\
                                      b_3
                                  \end{pmatrix}_{{\bf 3}}
 &= (a_1b_1+a_2b_2+a_3b_3)_{{\bf 1}} 
  \oplus \begin{pmatrix}
             \frac{1}{\sqrt 2}(a_2b_2-a_3b_3) \\                                            
             \frac{1}{\sqrt 6}(-2a_1b_1+a_2b_2+a_3b_3)
         \end{pmatrix}_{\bf 2} \nonumber \\
 &\ \oplus \begin{pmatrix}
            a_2b_3+a_3b_2 \\
            a_1b_3+a_3b_1 \\
            a_1b_2+a_2b_1
         \end{pmatrix}_{{\bf 3}} \oplus \begin{pmatrix}
                                          a_3b_2-a_2b_3 \\
                                          a_1b_3-a_3b_1 \\
                                          a_2b_1-a_1b_2
                                       \end{pmatrix}_{{\bf 3}'\ ,} \\
\begin{pmatrix}
a_1 \\
a_2 \\
a_3
\end{pmatrix}_{{\bf 3}'} \otimes  \begin{pmatrix}
                                      b_1 \\
                                      b_2 \\
                                      b_3
                                  \end{pmatrix}_{{\bf 3}'}
 &= (a_1b_1+a_2b_2+a_3b_3)_{{\bf 1}}
  \oplus \begin{pmatrix}
             \frac{1}{\sqrt 2}(a_2b_2-a_3b_3) \\                                            
             \frac{1}{\sqrt 6}(-2a_1b_1+a_2b_2+a_3b_3)
         \end{pmatrix}_{\bf 2} \nonumber \\
 &\ \oplus \begin{pmatrix}
            a_2b_3+a_3b_2 \\
            a_1b_3+a_3b_1 \\
            a_1b_2+a_2b_1
         \end{pmatrix}_{{\bf 3}} \oplus \begin{pmatrix}
                                          a_3b_2-a_2b_3 \\
                                          a_1b_3-a_3b_1 \\
                                          a_2b_1-a_1b_2
                                       \end{pmatrix}_{{\bf 3}'\ ,} \\
\begin{pmatrix}
a_1 \\
a_2 \\
a_3
\end{pmatrix}_{{\bf 3}} \otimes  \begin{pmatrix}
                                      b_1 \\
                                      b_2 \\
                                      b_3
                                  \end{pmatrix}_{{\bf 3}'}
 &= (a_1b_1+a_2b_2+a_3b_3)_{{\bf 1}'}  
 \oplus \begin{pmatrix}
             \frac{1}{\sqrt 6}(2a_1b_1-a_2b_2-a_3b_3) \\
             \frac{1}{\sqrt 2}(a_2b_2-a_3b_3)
         \end{pmatrix}_{\bf 2} \nonumber \\
 &\ \oplus \begin{pmatrix}
            a_3b_2-a_2b_3 \\
            a_1b_3-a_3b_1 \\
            a_2b_1-a_1b_2
         \end{pmatrix}_{{\bf 3}} \oplus \begin{pmatrix}
                                          a_2b_3+a_3b_2 \\
                                          a_1b_3+a_3b_1 \\
                                          a_1b_2+a_2b_1
                                       \end{pmatrix}_{{\bf 3}'\ .}
\end{align}
More details are shown in the review~\cite{Ishimori:2010au}.

\section{Next-to-leading order}

Parameters appeared in the down-type quark mass matrix with 
next-to-leading order are $\bar \epsilon_{ij}$. 
These are explicitly written as
\begin{align}
\bar \epsilon_{11}&=y_{\Delta _b}a _5a _{14}v_d
   +\bar y_{\Delta _{c_2}}a _1a _5v_d , \nonumber \\
\bar \epsilon_{12}&=
   -\frac{1}{2}y_{\Delta _b}a _5a _{14} v_d   
   -\left [  \frac{\sqrt 3}{4}(\sqrt 3-1)\bar y_{\Delta _{c_1}}-\frac{1}{4}(\sqrt 3+1)\bar y_{\Delta _{c_2}}  
   \right ]a _1a _5v_d , \nonumber \\
\bar \epsilon_{13}&=\left [\left \{ \frac{\sqrt 3}{4}(\sqrt 3-1)y_{\Delta _{a_1}}+\frac{1}{4}(\sqrt 3+1)y_{\Delta _{a_2}}\right \} a _1a _{13} 
   -\frac{1}{2}y_{\Delta _b}a _5a _{14}\right ]v_d  \nonumber \\
   &\ +\left [\left \{ -\frac{\sqrt 3}{4}(\sqrt 3+1)\bar y_{\Delta _{c_1}}-\frac{1}{4}(\sqrt 3-1)\bar y_{\Delta _{c_2}}\right \}a _1a _5
   +\frac{\sqrt 3}{2}\bar y_{\Delta _d}a _{13}a _{14}\right ]v_d , \nonumber \\
\bar \epsilon_{21}&=\bar y_{\Delta _{c_1}}a _1a _5v_d , \nonumber \\
\bar \epsilon_{22}&=\frac{\sqrt 3}{2}y_{\Delta _b}a _5a _{14} v_d
   -\left [ \frac{1}{4}(\sqrt 3-1)\bar y_{\Delta _{c_1}}
   +\frac{\sqrt 3}{4}(\sqrt 3+1)\bar y_{\Delta _{c_2}}  
   \right ]a _1a _5 v_d , \nonumber \\
\bar \epsilon_{23}&=\left [\left \{ -\frac{1}{4}(\sqrt 3-1)y_{\Delta _{a_1}}+\frac{\sqrt 3}{4}(\sqrt 3+1)
    y_{\Delta _{a_2}}\right \} a _1a _{13}
   -\frac{\sqrt 3}{2}y_{\Delta _b}a _5a _{14}\right ]v_d  \nonumber \\
   &\ +\left [\left \{ \frac{1}{4}(\sqrt 3+1)\bar y_{\Delta _{c_1}}
   -\frac{\sqrt 3}{4}(\sqrt 3-1)\bar y_{\Delta _{c_2}}\right \} a _1a _5
   -\frac{1}{2}\bar y_{\Delta _d}a _{13}a _{14}\right ]v_d , \nonumber \\
\bar \epsilon_{31}&=-y_{\Delta _e}a _5a _9v_d+\bar y_{\Delta _f}a _9a _{13}v_d , \nonumber \\
\bar \epsilon_{33}&=y_{\Delta _e}a _5a _9v_d .
\label{correction}
\end{align}

\section{Lepton sector}

The mass matrix of charged lepton becomes 
\begin{equation}
M_l = \begin{pmatrix}
                                   0 & -3y_1\lambda  a _9v_{45}/\sqrt 2 & 0 \\
                                   0 & -3y_1\lambda  a _9v_{45}/\sqrt 6 & 0 \\
                                   0 & 0 & y_2a _{13}v_d
                                \end{pmatrix},
\end{equation}
then, masses are given as
\begin{align}
m_e^2 = 0 \ ,
\quad
m_\mu ^2 =6|\bar y_1\lambda  a _9|^2v_d^2\ ,
\quad 
m_\tau ^2=|y_2|^2a _{13}^2v_d^2\ .
\label{chargemass}
\end{align}
In the same way, 
the right-handed Majorana mass matrix of neutrinos is given by 
\begin{equation}
M_N = \begin{pmatrix}
               y_1^N\lambda ^2\bar \Lambda +y_2^Na _4\Lambda & 0 & 0 \\
               0 & y_1^N\lambda ^2\bar \Lambda -y_2^Na _4\Lambda & 0 \\
               0 & 0 & M
         \end{pmatrix},
\end{equation}
and the Dirac mass matrix of neutrinos is
\begin{equation}
M_D = y_1^D\lambda v_u\begin{pmatrix}
        2a _5/\sqrt 6 & -a _5/\sqrt 6 & -a _5/\sqrt 6 \\
           0 & a _5/\sqrt 2 & -a _5/\sqrt 2 \\
                                        0 & 0 & 0 
                                     \end{pmatrix}+y_2^Dv_u\begin{pmatrix}
                                     0 & 0 & 0 \\
                                     0 & 0 & 0 \\
                         a _5 & a _5 & a _5
                                                \end{pmatrix}.
\end{equation}
By using the seesaw mechanism $M_\nu = M_D^TM_N^{-1}M_D$, 
the left-handed Majorana neutrino mass matrix is  written as
\begin{equation}
M_\nu = \begin{pmatrix}
                 a+\frac{2}{3}b & a-\frac{1}{3}b & a-\frac{1}{3}b \\
 a-\frac{1}{3}b& a+\frac{1}{6}b+\frac{1}{2}c & a+\frac{1}{6}b-\frac{1}{2}c \\
 a-\frac{1}{3}b & a+\frac{1}{6}b-\frac{1}{2}c & a+\frac{1}{6}b+\frac{1}{2}c
            \end{pmatrix},
\label{neutrino}
\end{equation}
where
\begin{equation}
a = \frac{(y_2^Da _5v_u)^2}{M},\qquad 
b = \frac{(y_1^Da _5v_u\lambda)^2}{y_1^N\lambda ^2\bar \Lambda +y_2^Na _4\Lambda},\qquad 
c = \frac{(y_1^Da _5v_u\lambda)^2}{y_1^N\lambda ^2\bar \Lambda -y_2^Na _4\Lambda}.
\label{neutrinomassparameter}
\end{equation}
It gives the tri-bimaximal mixing matrix 
$U_\text{tri-bi}$ and mass eigenvalues  as follows:
\begin{eqnarray}
&&U_\text{tri-bi} = \begin{pmatrix}
               \frac{2}{\sqrt{6}} &  \frac{1}{\sqrt{3}} & 0 \\
     -\frac{1}{\sqrt{6}} & \frac{1}{\sqrt{3}} &  -\frac{1}{\sqrt{2}} \\
      -\frac{1}{\sqrt{6}} &  \frac{1}{\sqrt{3}} &   \frac{1}{\sqrt{2}}
         \end{pmatrix},
\nonumber\\
\nonumber\\
&& m_{\nu_1} = b\ ,\qquad m_{\nu_2} = 3a\ ,\qquad m_{\nu_3} = c\ .
\label{mass123}
\end{eqnarray}
 The next-to-leading terms of the superpotential are important
to predict the deviation from the tri-bimaximal mixing of leptons.
 The relevant superpotential in the charged lepton sector
 is given at the next-to-leading order  as 
\begin{align}
\Delta w_l&=y_{\Delta _a}(T_1,T_2)\otimes (F_1,F_2,F_3)\otimes (\chi _1,\chi _2)\otimes (\chi _{11},\chi _{12},\chi _{13})\otimes H_{\bar 5}/\Lambda ^2 \nonumber \\
&\ +y_{\Delta _b}(T_1,T_2)\otimes (F_1,F_2,F_3)\otimes (\chi _5,\chi _6,\chi _7)\otimes \chi _{14}\otimes H_{\bar 5}/\Lambda ^2 \nonumber \\
&\ +y_{\Delta _c}(T_1,T_2)\otimes (F_1,F_2,F_3)\otimes (\chi _1,\chi _2)\otimes (\chi _5,\chi _6,\chi _7)\otimes H_{45}/\Lambda ^2 \nonumber \\
&\ +y_{\Delta _d}(T_1,T_2)\otimes (F_1,F_2,F_3)\otimes (\chi _{11},\chi _{12},\chi _{13})\otimes \chi _{14}\otimes H_{45}/\Lambda ^2 \nonumber \\
&\ +y_{\Delta _e}T_3\otimes (F_1,F_2,F_3)\otimes (\chi _5,\chi _6,\chi _7)\otimes (\chi _8,\chi _9,\chi _{10})\otimes H_{\bar 5}\otimes /\Lambda ^2 \nonumber \\
&\ +y_{\Delta _f}T_3\otimes (F_1,F_2,F_3)\otimes (\chi _8,\chi _9,\chi _{10})\otimes (\chi _{11},\chi _{12},\chi _{13})\otimes H_{45}\otimes /\Lambda ^2\ .
\label{nextsusy}
\end{align} 
By using this superpotential,
 we obtain the charged lepton mass matrix as
\begin{equation}
M_l\simeq
\begin{pmatrix}
\epsilon _{11} & \frac{\sqrt 3m_\mu}{ 2}+\epsilon_{12} & \epsilon _{13} \\
\epsilon _{21} & \frac{m_\mu}{2}+\epsilon_{22} & \epsilon _{23} \\
\epsilon _{31} & 0 & m_\tau+\epsilon_{33}
\end{pmatrix},
\label{nextleading}
\end{equation}
where $m_\mu$ and $m_\tau$ are given in  Eq.~(\ref{chargemass}) and
 $\epsilon_{ij}$'s are given as relevant linear combinations of 
$a_k a_l$'s. 
The explicit forms of   $\epsilon_{ij}$'s are given
by replacing  $\bar y_{\Delta_i}/3$ with  $-\bar y_{\Delta_i}$
 in $\bar \epsilon_{ij}$, which are presented in  Appendix B.
The charged lepton is diagonalized by 
the left-handed mixing matrix $U_E$ and the right-handed one $V_E$  as
\begin{eqnarray}
V_E^\dagger M_\ell U_E=M_\ell^\text{diag},
\end{eqnarray}
where $M_\ell^\text{diag}$ is a diagonal matrix.
These mixing matrices can be written by
\begin{eqnarray}
\label{VEUE}
\begin{split}
V_E&=
\begin{pmatrix}
       \cos 60^\circ &  \sin 60^\circ &  0\\
    -\sin 60^\circ & \cos 60^\circ  &  0\\
    0& 0& 1
\end{pmatrix}
\times
\begin{pmatrix}
       1 &  \frac{\tilde a^2}{\lambda^2} &  \tilde a\\
    -\frac{\tilde a^2}{\lambda^2}-\tilde a^2  & 1  &  \tilde a\\
    -\tilde a+ \frac{\tilde a^3}{\lambda^2} 
& -\tilde a-  \frac{\tilde a^3}{\lambda^2}& 1
\end{pmatrix},
\\
U_E&=
 \begin{pmatrix}
       1 &  \frac{\tilde a}{\lambda} &  \tilde a\\
    -\frac{\tilde a}{\lambda}-\tilde a^2  & 1  &  \tilde a\\
    -\tilde a+ \frac{\tilde a^2}{\lambda} 
& -\tilde a-  \frac{\tilde a^2}{\lambda}& 1
\end{pmatrix}.
\end{split}
\end{eqnarray}
Taking the next-to-leading order, the electron has non-zero mass, namely
\begin{eqnarray}
m_e^2 &\simeq 
\frac{3}{2}\left (\frac{1}{6}\epsilon _{11}^2
-\frac{1}{\sqrt 3}\epsilon _{11}\epsilon _{21}+\frac{1}{2}\epsilon _{21}^2\right )\simeq {\cal O}(\tilde a^4 v_d^2).
\end{eqnarray}

\section{Formulae for quark sector}
\label{sec:formulae}
Here we will give formulae for quark sector which are used in our analysis. 
The SUSY contribution by gluino-squark box diagram to the dispersive part of 
the effective Hamiltonian for $M-\bar M$ mixing ($M=K,B_d,B_s$) is given by \cite{Gabbiani:1996hi,Altmannshofer} 
\be
M_{12}^{M,\text{SUSY}} &=& -\frac{\alpha _S^2}{216m_{\tilde q}^2}\frac{2}{3}M_M f_M^2
\Bigg [ \left \{ (\delta_d^{LL})_{ij}^2+(\delta_d^{RR})_{ij}^2\right \} 
\left \{ 24xf_6(x)+66\tilde f_6(x)\right \} \nonumber \\
&+&(\delta _d^{LL})_{ij}(\delta _d^{RR})_{ij}\left(\left \{ 384\left (\frac{M_M}{m_j+m_i}\right )^2+72\right \} xf_6(x) 
+\left \{ -24\left (\frac{M_M}{m_j+m_i}\right )^2+36\right \} \tilde f_6(x)\right) \nonumber \\
&+&\left \{ (\delta _d^{LR})_{ij}^2+(\delta _d^{RL})_{ij}^2\right \} \left \{ -132\left (\frac{M_M}{m_j+m_i}\right )^2\right \} xf_6(x) \nonumber \\
&+& (\delta _d^{LR})_{ij}(\delta _d^{RL})_{ij} \left \{ -144\left (\frac{M_M}{m_j+m_i}\right )^2-84\right \} \tilde f_6(x)\Bigg ],
\ee
where $x=m_{\tilde g}^2/m_{\tilde q}^2$ and the loop functions are defined as
\be
f_6(x)&=&\frac{6(1+3x)\log x+x^3-9x^2-9x+17}{6(x-1)^5}, \\
\tilde f_6(x)&=&\frac{6x(1+x)\log x-x^3-9x^2+9x+1}{3(x-1)^5}.
\ee
For $M=K,B_d,B_s$ meson system, the generation indices of down-type quarks $(i,j)$ correspond to $(i,j)=(1,2),(1,3),(2,3)$, respectively. 

For $b \to s \gamma$ decay, 
the Branching Ratio (BR) is given by 
\be
\text{BR}(b\rightarrow s\gamma )=\alpha _s^2\alpha \frac{m_b^3 \tau_B }{81\pi ^2m_{\tilde q}^4}
\left [\left |m_b G_3(x)(\delta _d^{LL})_{23}+m_{\tilde g}G_1(x)(\delta _d^{LR})_{23}\right |^2
+(L \leftrightarrow R)\right], 
\ee 
where $\tau_B$ is the lifetime of the B meson, and the loop functions are defined as 
\be
G_1(x)&=&\frac{1+4x-5x^2+4x\log x+2x^2\log x}{2(x-1)^4}, \\
G_3(x)&=&\frac{-1+9x+9x^2-17x^3+18x^2\log x +6x^3\log x}{12(x-1)^5}.
\ee
The chromo EDM of the strange quark is given by \cite{Hisano:2003iw}
\be
d_s^C=c\frac{\alpha _s}{4\pi }\frac{m_{\tilde g}}{m_{\tilde q}^2}\left (-\frac{1}{3}N_1(x)-3N_2(x)\right )
\text{Im}[(\delta _d^{LL})_{23}(\delta _d^{LR})_{33}(\delta _d^{RR})_{32}], 
\ee
where $c$ is the QCD correction. We take $c=0.9$. The functions $N_1(x)$ and $N_2(x)$ are given as follows:
\be
N_1(x)&=&\frac{3+44x-36x^2-12x^3+x^4+12x(2+3x)\log x}{6(x-1)^6}, \\
N_2(x)&=&-\frac{10+9x-18x^2-x^3+3(1+6x+3x^2)\log x}{3(x-1)^6}.
\ee

\section{$\mu\to e\gamma$, $\tau\to e\gamma$ and $\tau\to \mu\gamma$}

In the framework of SUSY, LFV effects  originate 
from  misalignment between fermion and sfermion mass eigenstates.
 Once non-vanishing off-diagonal elements of the slepton mass matrices
are generated in the super-CKM basis,
LFV rare decays like $\ell_i\to\ell_j\gamma$ are naturally induced by one-loop diagrams with the exchange of gauginos 
and sleptons. 
The decay $\ell_i\to\ell_j\gamma$ is described by the dipole operator and the corresponding amplitude reads
\cite{Gabbiani:1996hi,Hisano:1995cp,Borzumati:1986qx,Hisano:1995nq,Hisano:2009ae}
\begin{eqnarray}
T=m_{\ell_i}\epsilon^{\lambda}\overline{u}_j(p-q)[iq^\nu\sigma_{\lambda\nu}
(A_{L}P_{L}+A_{R}P_{R})]u_i(p)\,,
\end{eqnarray}
where $p$ and $q$ are momenta of the initial lepton $\ell_i$ and of the photon, respectively, 
and $A_{L,R}$ are the two possible amplitudes in this  process. 
The branching ratio of $\ell_{i}\rightarrow \ell_{j}\gamma$ can be written 
as follows:
\begin{eqnarray}
\frac{{\rm BR}(\ell_{i}\rightarrow  \ell_{j}\gamma)}{{\rm BR}(\ell_{i}\rightarrow 
\ell_{j}\nu_i\bar{\nu_j})} =
\frac{48\pi^{3}\alpha}{G_{F}^{2}}(|A_L^{ij}|^2+|A_R^{ij}|^2)\,.
\nonumber
\end{eqnarray}
In the mass insertion approximation, it is found that~\cite{Altmannshofer}
\begin{eqnarray}
\label{MIamplL}
A^{ij}_L
&\simeq&\frac{\alpha_2}{4\pi}
\frac{\left(\delta^{LL}_{\ell}\right)_{ij}}{m_{\tilde \ell}^{2}}\tan{\beta}
~\bigg[
\frac{\mu M_{2}}{(M_{2}^2-\mu^2)}\bigg(f_{2n}(x_2,x_\mu)+f_{2c}(x_2,x_\mu)\bigg)
\nonumber\\
&&+ \tan^2\theta_{W}\,
\mu M_{1}\bigg(\frac{f_{3n}(x_1)}{m_{\tilde \ell}^{2}}+
\frac{f_{2n}(x_1,x_\mu)}{(\mu^2 - M_{1}^2)}\bigg)
\bigg]
+ \frac{\alpha_1}{4\pi}~\frac{\left(\delta^{RL}_{\ell}\right)_{ij}}{m_{\tilde \ell}^2}~
\left(\frac{M_1}{m_{\ell_i}}\right)~2~f_{2n}(x_1)~,
\nonumber\\
A^{ij}_R
&\simeq&
\frac{\alpha_{1}}{4\pi}
\left[
\frac{\left(\delta^{RR}_{e}\right)_{ij}}{m_{\tilde \ell}^{2}}\mu M_{1}\tan{\beta}
\left(\frac{f_{3n}(x_1)}{m_{\tilde \ell}^{2}}-\frac{2f_{2n}(x_1,x_{\mu})}{(\mu^2-M_{1}^2)}\right)
+2\frac{\left(\delta^{LR}_{e}\right)_{ij}}{m_{\tilde\ell}^{2}}~
\left(\frac{M_1}{m_{\ell_i}}\right)~f_{2n}(x_1)
\right]~,
\nonumber\\
\label{form}
\end{eqnarray}
where $\theta_W$ is the weak mixing angle, 
$x_{1,2}=M_{1,2}^2/m_{\tilde \ell}^2$, $x_\mu=\mu^2/m_{\tilde \ell}^2$ 
and $f_{i(c,n)}(x,y)=f_{i(c,n)}(x)-f_{i(c,n)}(y)$. The loop functions 
$f_i$'s are given explicitly as follows:
 \begin{eqnarray}
\begin{split}
f_{2n}(x) &= \frac{-5x^2+4x+1+2x(x+2)\log x}{4(1-x)^4}~, 
\\
f_{3n}(x) &= \frac{1+9x-9x^2-x^3+6x(x+1)\log x}{3(1-x)^5}~,
\\
f_{2c}(x) &= \frac{-x^2-4x+5+2(2x+1)\log x}{2(1-x)^4}~.
\end{split}
\end{eqnarray}

\section{Electron electric dipole moment}

The mass insertion parameters also contribute to the electron EDM 
through one-loop exchange of binos/sleptons. 
The corresponding EDM is given as
\cite{Hisano:2007cz,Hisano:2008hn,Altmannshofer}
\begin{eqnarray}
\label{Eq:lEDM_LO}
\frac{d_{e}}{e}
\!\!=\!\!
-\frac{\alpha_1}{4\pi}\frac{M_1}{m^{2}_{\tilde\ell}}\!\!\!
&\bigg\{&
\!\!\!{\rm Im}
[(\delta^{LR}_{\ell})_{1k}(\delta^{RR}_{e})_{k1} + (\delta^{LL}_{\ell})_{1k}(\delta^{LR}_{\ell})_{k1}]
\,f_{3n}(x_1)
+{\rm Im}[(\delta^{LL}_{\ell})_{1k}(\delta^{LR}_{\ell})_{kl}(\delta^{RR}_{e})_{l1}
\nonumber\\
&+&
(\delta^{LR}_{\ell})_{1k}(\delta^{RR}_{e})_{kl}(\delta^{RR}_{e})_{l1}+
(\delta^{LL}_{\ell})_{1k}(\delta^{LL}_{\ell})_{kl}(\delta^{LR}_{\ell})_{l1}]
\,f_{4n}(x_1)
\bigg\}\,,
\end{eqnarray}
where $k,l=2,3$, $(\delta^{LR}_{\ell})_{33}= -m_{\tau}(A_{\tau}+\mu\tan\beta)/m^{2}_{\tilde\ell}$, and  the loop function $f_{4n}$
 is given as 
\begin{eqnarray}
\begin{split}
f_{4n}(x) &=   \frac{-3-44x+36x^2+12x^3-x^4-12x(3x+2)\log x}{6(1-x)^6}\, .
\end{split}
\end{eqnarray}

Since components $(i,3)$ and $(3,i)$ of $\delta_e^{RR}$ are 
much larger compared to others in our model, 
dominant terms are given as
\begin{eqnarray}
\label{Eq:lEDM_LO}
\frac{d_{e}}{e}
\approx
-\frac{\alpha_1}{4\pi}\frac{M_1}{m^{2}_{\tilde\ell}}
&\bigg\{&
\mathcal{O}(\frac{m_e}{m_{\tilde\ell}}a_1)
\,f_{3n}(x_1)
+\mathcal{O}(\frac{m_\tau}{m_{\tilde\ell}}(1+\frac{\mu\tan\beta}{m_{\tilde\ell}})
a_1\tilde a^2)
\,f_{4n}(x_1)
\bigg\}.
\end{eqnarray}


	
\bibliographystyle{unsrt}

\end{document}